\documentclass[onecolumn]{aastex61}\usepackage{amsmath, amsthm,amssymb}\usepackage{natbib}
\usepackage{color,xcolor,colortbl}
\makeatletter
\newcommand{\ssymbol}[1]{^{\@fnsymbol{#1}}}
\makeatother

\begin{document}
\title{Col-OSSOS: Color and Inclination are Correlated Throughout the Kuiper Belt}

\author[0000-0001-8617-2425]{Micha$\ddot{\rm e}$l Marsset}
\affiliation{Astrophysics Research Centre, Queen's University Belfast, Belfast BT7 1NN, United Kingdom}
\affiliation{Department of Earth, Atmospheric and Planetary Sciences, MIT, 77 Massachusetts Avenue, Cambridge, MA 02139, USA}

\author[0000-0001-6680-6558]{Wesley C. Fraser}
\affiliation{Astrophysics Research Centre, Queen's University Belfast, Belfast BT7 1NN, United Kingdom}

\author[0000-0003-4797-5262]{Rosemary E. Pike}
\affiliation{Institute of Astronomy and Astrophysics, Academia Sinica; 11F of AS/NTU Astronomy-Mathematics Building, No.1, Sec. 4, Roosevelt Rd, Taipei 10617, Taiwan, R.O.C.}

\author[0000-0003-3257-4490]{ Michele T. Bannister}
\affiliation{Astrophysics Research Centre, Queen's University Belfast, Belfast BT7 1NN, United Kingdom}
\affiliation{Department of Physics and Astronomy, University of Victoria, Elliott Building, 3800 Finnerty Rd, Victoria, BC V8P 5C2, Canada}
\affiliation{NRC-Herzberg Astronomy and Astrophysics, National Research Council of Canada, 5071 West Saanich Rd, Victoria, BC V9E 2E7, Canada}

\author[0000-0003-4365-1455]{Megan E. Schwamb}
\affiliation{Gemini Observatory, Northern Operations Center, 670 North A'ohoku Place, Hilo, HI 96720, USA}

\author[0000-0001-8736-236X]{Kathryn Volk}
\affiliation{Lunar and Planetary Laboratory, The University of Arizona, 1629 E. University Blvd., Tucson, AZ 85721, USA}

\author[0000-0001-7032-5255]{J. J. Kavelaars}
\affiliation{NRC-Herzberg Astronomy and Astrophysics, National Research Council of Canada, 5071 West Saanich Rd, Victoria, BC V9E 2E7, Canada}
\affiliation{Department of Physics and Astronomy, University of Victoria, Elliott Building, 3800 Finnerty Rd, Victoria, BC V8P 5C2, Canada}

\author[0000-0003-4143-8589]{Mike Alexandersen} 
\affiliation{Institute of Astronomy and Astrophysics, Academia Sinica; 11F of AS/NTU Astronomy-Mathematics Building, No.1, Sec. 4, Roosevelt Rd, Taipei 10617, Taiwan, R.O.C.}

\author[0000-0001-7244-6069]{Ying-Tung Chen} 
\affiliation{Institute of Astronomy and Astrophysics, Academia Sinica; 11F of AS/NTU Astronomy-Mathematics Building, No.1, Sec. 4, Roosevelt Rd, Taipei 10617, Taiwan, R.O.C.}

\author{Brett J. Gladman}
\affiliation{Department of Physics and Astronomy, University of British Columbia, Vancouver, BC V6T 1Z1, Canada}

\author{Stephen D. J. Gwyn}
\affiliation{NRC-Herzberg Astronomy and Astrophysics, National Research Council of Canada, 5071 West Saanich Rd, Victoria, BC V9E 2E7, Canada}

\author[0000-0003-4077-0985]{Matthew J. Lehner}
\affiliation{Institute of Astronomy and Astrophysics, Academia Sinica; 11F of AS/NTU Astronomy-Mathematics Building, No.1, Sec. 4, Roosevelt Rd, Taipei 10617, Taiwan, R.O.C.}
\affiliation{Department of Physics and Astronomy, University of Pennsylvania, 209 S. 33rd St., Philadelphia, PA 19104, USA}
\affiliation{Harvard-Smithsonian Center for Astrophysics, 60 Garden St., Cambridge, MA 02138, USA}

\author[0000-0002-6830-476X]{Nuno Peixinho}
\affiliation{CITEUC -- Centre for Earth and Space Science Research of the University of Coimbra, Geophysical and Astronomical Observatory of the University of Coimbra, 3040-004 Coimbra, Portugal.}

\author[0000-0003-0407-2266]{Jean-Marc Petit}
\affiliation{Institut UTINAM UMR6213, CNRS, Univ. Bourgogne Franche-Comt\'e, OSU Theta F25000 Besan\c{c}on, France}

\author{Shiang-Yu Wang}
\affiliation{Institute of Astronomy and Astrophysics, Academia Sinica; 11F of AS/NTU Astronomy-Mathematics Building, No.1, Sec. 4, Roosevelt Rd, Taipei 10617, Taiwan, R.O.C.}

\email{michael.marsset@qub.ac.uk}

\begin{abstract}

Both physical and dynamical properties must be considered to constrain the origins of the dynamically excited distant Solar System populations. 
We present high-precision (\emph{g}-\emph{r}) colors for 25 small (H$_{\rm r}$$>$5) dynamically excited Trans-Neptunian Objects (TNOs) and centaurs acquired as part of the Colours of the Outer Solar System Origins Survey (Col-OSSOS). 
We combine our dataset with previously published measurements and consider a set of 229 colors of outer Solar System objects on dynamically excited orbits. 
The overall color distribution is bimodal and can be decomposed into two distinct classes, termed `gray' and `red', that each has a normal color distribution. 
The two color classes have different inclination distributions: red objects have lower inclinations than the gray ones. 
This trend holds for all dynamically excited TNO populations. 
Even in the worst-case scenario, biases in the discovery surveys cannot account for this trend: it is intrinsic to the TNO population. 
Considering that TNOs are the precursors of centaurs, and that their inclinations are roughly preserved as they become centaurs, our finding solves the conundrum of centaurs being the only outer Solar System population identified so far to exhibit this property \citep{Tegler:2016ep}. 
The different orbital distributions of the gray and red dynamically excited TNOs provide strong evidence that their colors are due to different formation locations in a disk of planetesimals with a compositional gradient. 

\end{abstract}

\keywords {Kuiper belt: general  -- minor planets, asteroids: general  -- surveys}



\section{Introduction}
\label{sec:intro}

Trans-Neptunian Objects (TNOs) represent some of the most unaltered remnants of the planetary formation process. The surface of these objects can be used as tracers of their dynamical evolution which, in turn, helps us understand the current structure of the Kuiper belt (e.g., \citealt{Gladman:2005hw, Petit:2011hj}). 
Numerical models have been used to assess the evolution of TNOs from their formation locations onto their current orbits (e.g., \citealt{Malhotra:1995dy, Gomes:2003cd, Levison:2008ib, Nesvorny:2015et, Nesvorny:2015ik}). 
Different formation locations might correspond to distinct surface compositions, though this remains disputed \citep{GilHutton:2002bh, Trujillo:2002ew, Stern:2002bj, Delsanti:2004jg, SantosSanz:2009gq}. 
As such, both physical and dynamical properties must be considered to constrain the origins of TNOs. 
Unlike the majority of the known dwarf-planet sized bodies, the smaller TNOs are too faint to be studied through optical and infrared spectroscopy. For these objects, one must instead rely on what broad-band colors reveal. 

The TNOs can be divided into two broad dynamical groups of objects: the dynamically quiescent `cold classicals', and the dynamically excited `hot population' \citep{Tegler:2000fx, Brown:2001fu}. 
The latter comprises several sub-populations: 
1) the hot classicals located at 40-48~astronomical units (au) between Neptune's 3:2 and 2:1 mean-motion orbital resonances (MMRs; \citealt{Levison:2001fc, Doressoundiram:2002jw, Gladman:2008tu}), 2) the resonant objects locked in MMRs with Neptune \citep{Malhotra:1995dy, Levison:2008ib, Gladman:2012ed}, 3) the scattering population on dynamically unstable and highly eccentric orbits \citep{Lykawka:2007kc, Gladman:2008tu, Gomes:2008ud} and, 4) the detached TNOs that experienced gravitational excitation in the past and are now on stable orbits unperturbed by Neptune \citep{Gladman:2002dq, Emelyanenko:2003gw, Gomes:2005fl, Brasser:2015iw}. Centaurs are thought to be former TNOs that evolved into inner orbits with semi-major axes and perihelia between the orbits of Jupiter and Neptune (e.g., \citealt{Levison:1997el}). We consider them here as part of the larger group of dynamically excited TNOs.

There are indications that colors and orbital properties of the dynamically excited TNOs are related. 
The largest dynamically excited TNOs uniformly exhibit neutral, ice-dominated surfaces \citep{Fornasier:2004io, Barkume:2008gw, Barucci:2008wx, Barucci:2011bd, Brown:2012cv}, possibly because they have sufficient gravity to retain their volatiles \citep{Schaller:2007cq}. 
In contrast, smaller dynamically excited TNOs have a bimodal color distribution, with classes termed `gray' and `red' that are divided around optical color (\emph{g}-\emph{r})=0.78 \citep{Tegler:1998jl, Tegler:2000fx, Tegler:2003jx, Tegler:2003is, Tegler:2008vj, Tegler:2016ep, Peixinho:2003jn, Peixinho:2012bt, Peixinho:2015bw, Fraser:2012cs, Lacerda:2014hw, Wong:2017ck}; this is fully quantified in section~\ref{sec:results}. 
A correlation between colors and orbital inclinations was found for the population of classical TNOs \citep{Tegler:2000fx, Trujillo:2002ew, Hainaut:2002fx, Doressoundiram:2005ha, Peixinho:2008gn}. 
However, the datasets in these works were contaminated by the presence in their sample of peculiar compositional groups, including neutral-color Haumea family members \citep{Brown:2007kh, Ragozzine:2007jx, Schaller:2008is, Snodgrass:2010cf} and red cold classical objects \citep{Tegler:2000fx, Brown:2001fu}. 
A careful removal of these objects diminishes the strength of the correlation between color and inclination, but the correlation remains statistically significant \citep{Peixinho:2015bw}. 
More recently, \citet{Tegler:2016ep} divided the observed population of centaurs into two groups by color and analyzed their orbital inclination distributions. 
They found that red centaurs have lower inclinations than the gray ones. 
As minor planets on trans-neptunian orbits are perturbed into the centaur region, they roughly preserve the inclinations they had before becoming centaurs \citep{Volk:2013gv}. 
As such, the different orbital distributions of the two color classes of centaurs should be reflective of a similar trend in the more distant TNOs. 
Such a trend might be the origin of the observed color versus inclination correlation of the hot classical population. 
Two color groups with different inclination distributions could indeed manifest as such a correlation. 
In short, different inclinations for the gray and red dynamically hot objects are anticipated but not yet observed in the Kuiper belt.

We reexamine here the orbital inclination distribution of the two color classes of the dynamically excited TNOs. 
We use a large dataset of high-quality color measurements that includes optical (\emph{g}-\emph{r}) colors obtained for 44 objects as part of the Colours of the Outer Solar System Origin Survey \citep{Schwamb:2018}, 25 of which are new measurements (section~\ref{sec:obs}). 
We further use robust color measurements extracted from several works reporting optical colors of TNOs, bringing our sample to 229 dynamically excited TNOs. 
We show in section~\ref{sec:results} that red dynamically excited TNOs have smaller orbital inclinations than their gray counterparts and that this trend appears to be present in every dynamical population. 
The centaurs are no longer the only population to exhibit this peculiarity. 
This has strong implications for the origins of the dynamically excited TNO populations (section~\ref{sec:discussion}).

\section{Sample selection, observations, and color measurements}
\label{sec:obs}

\subsection{Sample selection}
\label{sec:sample}

Our analysis uses both newly obtained colors and color measurements from the literature for TNOs confirmed to be on dynamically excited orbits. 
We consider the subset of 44 dynamically excited objects from the Outer Solar System Origin Survey (OSSOS; \citealt{Bannister:2016cp, Bannister:2018ha}) for which our team acquired optical and near-infrared colors as part of the Colours of OSSOS (Col-OSSOS) survey \citep{Schwamb:2018}. 
Measurements for 19 of these objects are previously-published colors 
from \citet{Schwamb:2018}, and 
25 are new color measurements reported here.  
Those include measurements for four objects that we previously published in \citet{Pike:2017gf} and reprocessed here using the latest version of our photometric pipeline.
We complement this Col-OSSOS dataset with available high-quality color measurements from the literature. 
These include optical colors from \citet{Peixinho:2015bw}, which compiles data from many authors, and  \citet{Tegler:2016ep}, as well as HST data from \citet{Fraser:2012cs} and \citet{Fraser:2015cx}. 
Only colors observed with filters in the $\sim$500-850~nm range were considered, i.e., Johnson-Cousins (V-R) and (R-I) and HST (F606w-F814w). 
One object from \citet{Peixinho:2015bw}, 2002~GO9 (Crantor), has (R-I) color inconsistent with its (V-R) color and measured spectra (e.g., \citealt{AlvarezCandal:2007jg}). 
We consider only the (V-R) measurement reported in \citet{Peixinho:2015bw} for that target.

Only small objects are considered for this analysis. 
The size transition from ice-rich surfaces on large TNOs to potentially volatile depleted surfaces on small TNOs is a matter of debate, but the transition is generally proposed to be at absolute magnitude ${\rm H_r}$$\sim$6 \citep{Peixinho:2008gn, Fraser:2012cs}. Here, we find that including objects in the 5$<$${\rm H_r}$$<6$ magnitude range does not affect the conclusions of our analysis (see section~\ref{sec:results}). We therefore adopt a magnitude cut of ${\rm H_r}>$5.0, corresponding to diameters of $\lesssim$440\,km assuming a visible geometrical albedo of 0.09 \citep{Lacerda:2014hw}. 


We classified all the TNOs in our sample using the procedure outlined in \citet{Gladman:2008tu}, which we briefly review here. 
Each object's orbit was fit using the \citet{Bernstein:2000bk} orbit fitting routine. 
Starting from that best-fit orbit, a Monte-Carlo search was performed for the minimum and maximum semi-major axis orbits that do not produce orbit-fit residuals more than 1.5 times worse than those of the best fit orbit. 
All three orbits (the best fit, minimum-a, and maximum-a) were integrated forward in time for 10~Myr and classified according to their evolution. 
A secure classification is one for which all three orbits behave the same way, and insecure classification is one where at least one orbit behaves differently. 
We have 202 secure classifications and 27 insecure classifications. 
Most of the insecure classifications are a result of an object being in or near a MMR rather than having a particularly large uncertainty in semi-major axis because all of the TNOs in our dataset have reasonably long ($>$3 oppositions) observational arcs.

Objects belonging to compositionally distinct TNO populations, which we term `interlopers', were rejected from our sample. Specifically, we excluded all known Haumea family members, as well as gray-colored OSSOS/Col-OSSOS objects orbiting within the Haumea family cluster (confirmed members have $41.6<a<44.2$~au, $0.08<e<0.19$, $24.2<i<29.1\degr$; \citealt{Brown:2007kh, Ragozzine:2007jx, Schaller:2008is, Snodgrass:2010cf}). 
We further rejected cold classical objects, which are known for exhibiting different colors \citep{Tegler:2000fx, Pike:2017gf} and albedos \citep{Brucker:2009jp, Vilenius:2014jf} from those of the dynamically excited TNOs. 
\citet{Gulbis:2010fw} and \citet{Petit:2011hj} found that the orbital inclination distribution of cold classicals can be best reproduced by a Gaussian of width $2.0\degr^{+0.6}_{-0.5}$ or sin($i$) times a Gaussian of width $2.6\degr$, respectively. 
Adopting an inclination cut of $i>5\degr$ ensures contamination is minimal. 
The same inclination cut was applied to all objects in our dataset, independently of their dynamical classification, in order to perform a consistent analysis throughout all populations. We further verified that changing the inclination cut does not affect the conclusions of our analysis (section~\ref{sec:results}). Centaurs with Tisserand parameter T$_{\rm J}$$<$3 were also excluded from our sample as their orbits are strongly coupled to Jupiter (e.g., \citealt{Gladman:2008tu}) and thus have likely experienced larger orbital perturbations and changes in inclination (e.g., \citealt{Volk:2013gv}). 
The smallest (H$>$11) centaurs and scattering disk objects appear to have altered colors compared to the rest of the centaur and scattering population \citep{Jewitt:2015fi}, so they were also excluded. No other populations in our dataset contain objects with H$>$11. Finally, we also rejected the two known objects with retrograde orbits because their origin population remains elusive \citep{Gladman:2009kv, Chen:2016hr}. Orbital elements and dynamical classifications for the resulting dataset of 229 objects are reported in Table~\ref{tab:sptslopes} (appendix~\ref{sec:dataset}).

\subsection{Observations, color measurements and spectral slopes}

All new color measurements presented in this paper were acquired through the Col-OSSOS large program (PI: W. Fraser) on Gemini North between August 2014 and November 2016.
Col-OSSOS collects near-simultaneous \emph{g}, \emph{r}, and \emph{J} band photometry of a magnitude-limited (\emph{r}$<$23.6) subset of the OSSOS sample. 
Optical measurements were acquired with the Gemini Multi-Object Spectrograph (GMOS, \citealt{Hook:2004fq}), and the \emph{J} band sequence was obtained with the Near InfraRed Imager and Spectrometer (NIRI, \citealt{Hodapp:2003ko}); we observed using a \emph{rgJgr} sequence to account for any brightness variation due to lightcurve effects. 

A complete description of the data reduction and photometric extraction for Col-OSSOS is provided in \citet{Schwamb:2018}. We summarise it briefly here.
Processing of the Col-OSSOS data was performed with the Gemini-IRAF package. We first used the bias images acquired as part of the GMOS calibration plan to remove the bias offset from the science frames. Master sky flats were produced from a set of twilight flats and used to flatten the science frames. Photometric measurements were achieved using the TRIPPy software package, which makes use of a pill-shaped aperture, built from the convolution product of a circular aperture with a line describing the direction and rate of motion of the target \citep{Fraser:2016bd}. This approach minimizes the background contribution to the total flux in the photometric aperture.

Next, the Col-OSSOS data were calibrated to the SDSS magnitude catalog \citep{SDSSCollaboration:2016tq} in cases where the field of view of the target was covered by SDSS; in other cases, we calibrated to the Pan-STARRS magnitude catalog \citep{Magnier.2013, PanStarrsDR1:2016}. 
We used in-frame catalog stars to determine the color conversion between the SDSS or Pan-STARRS and GMOS filter sets, using the relationships provided in \citet{Schwamb:2018}. We then fit the calibrated \emph{g} and \emph{r} magnitudes with a linear lightcurve model assuming a constant (\emph{g}-\emph{r}) color across the observing sequence. The fit was weighted by the uncertainty of the individual photometric measurements calculated as the quadratic sum of the photometric uncertainty, the zeropoint uncertainty, and the aperture correction uncertainty. The (\emph{g}-\emph{r}) color was then derived from our best fit, and the uncertainty on the color computed as the uncertainty on the fit.
New colors obtained from our observations are reported in Table~\ref{tab:objects}.

All color measurements from Col-OSSOS and the literature were converted to spectral slope (s), defined as the percentage increase in reflectance per ${\rm 10^3 \AA}$ change in wavelength normalised to 550~nm, using the Synphot tool in the STSDAS software package\footnote{www.stsci.edu/institute/software\_hardware/stsdas}. 
Reported spectral slopes are the mean values from all the measurements of a target weighted by the inverse of the squared uncertainties. We consider only well-quantified measurements with spectral slope uncertainty ${\rm \Delta s<12\%/(10^3 \AA})$, because larger error bars do not allow clear differentiation between the two color classes. The full list of measurements we consider is provided in Table~\ref{tab:sptslopes} (appendix~\ref{sec:dataset}). 

\startlongtable
\begin{deluxetable}{llcccccccc}
\tabletypesize{ \scriptsize}
\tablecaption{\label{tab:objects} New Col-OSSOS color measurements and observing circumstances. }
\tablehead{\colhead{Target} & \colhead{MPC} &  \colhead{OSSOS} & \colhead{H${\rm _r^*}$} &  \colhead{Col-OSSOS} & \colhead{$\Delta$ (au)} & \colhead{r (au)} &  \colhead{$\alpha$} &  \colhead{Mean} \\ 
\colhead{Designation} & \colhead{Designation} & \colhead{r$^{\prime}$ mag$^*$} & \colhead{  } & \colhead{(\emph{g'}-\emph{r'})}  & \colhead{  } & \colhead{  } & \colhead{} &  \colhead{MJD} & }
\startdata
o3e01 & 2002GG166 & 21.50$\pm$0.09 & 7.73 &      0.63$\pm$0.01 & 21.69 & 20.70 & 0.52 & 57133.47513 \\
o3e02 & 2013GH137 & 23.34$\pm$0.14 & 8.32 &      0.77$\pm$0.03 & 31.45 & 30.46 & 0.29 & 57519.33308 \\
o3e05 & 2013GW136 & 22.69$\pm$0.07 & 7.42 &     0.78$\pm$0.01 & 32.62 & 31.65 & 0.46 & 57130.40010 \\
o3e09 & 2013GY136  & 22.94$\pm$0.05 & 7.32 &     0.55$\pm$0.01 & 35.22 & 34.24 & 0.38 & 57132.42388 \\
o3e23PD & 2001FO185 & 23.37$\pm$0.08 & 7.09 &  0.87$\pm$0.06 & 42.03 & 41.03 & 0.20 & 57516.79552 \\ 
o3e29 & 2013GO137 & 23.46$\pm$0.08 & 7.09 &      0.78$\pm$0.04 & 42.53 & 41.53 & 0.05 & 57139.41226 \\
o3e39  & 2013GP136  & 23.07$\pm$0.07 & 6.42 &     0.74$\pm$0.02 & 44.75 & 43.75 & 0.13 & 57133.96175 \\
o3e44 & 2013GG138 & 23.26$\pm$0.09 & 6.34 &      1.18$\pm$0.03 & 48.39 & 47.39 & 0.16 & 57135.40627 \\
o3o09  & 2013JB65 & 23.23$\pm$0.06 & 8.13 &         0.76$\pm$0.02 & 31.98 & 31.00 & 0.47 & 57518.49578 \\
o3o11(1)  & 2013JK64 & 22.94$\pm$0.05 & 7.69 &     1.11$\pm$0.01 & 33.43 & 32.48 & 0.61 & 57514.57625 \\
o3o11(2)  & 2013JK64 & 22.94$\pm$0.05 & 7.69 &     0.98$\pm$0.01 & 33.28 & 32.47 & 1.03  & 57130.63127 \\
o3o14  & 2013JO64 & 23.54$\pm$0.08 & 8.00 &         0.59$\pm$0.02 & 35.74 & 34.92 & 0.93 & 57130.50551 \\
o3o15 & 2013JD65  & 23.48$\pm$0.07 & 7.90 &         0.83$\pm$0.03 & 35.70 & 34.85 & 0.88 & 57132.55296 \\
o3o18  & 2013JE64 & 23.56$\pm$0.15 & 7.94 &         0.41$\pm$0.13 & 36.65 & 35.66 & 0.36 & 57519.50134 \\
o3o20PD & 2007JF43 & 21.15$\pm$0.02 & 5.27 &     1.01$\pm$0.01 & 37.96 & 37.08 & 0.74 & 57135.50416 \\
o3o21  & 2013JR65 & 23.51$\pm$0.12 & 7.53 &         0.46$\pm$0.03 & 38.83 & 37.86 & 0.47 & 57550.43376 \\
o3o27  & 2013JJ65 & 23.40$\pm$0.08 & 7.22 &          1.10$\pm$0.03 & 41.43 & 40.62 & 0.82 & 57131.54637 \\
o3o28  & 2013JN65 & 23.42$\pm$0.21 & 7.23 &         0.57$\pm$0.01 & 41.17 & 40.20 & 0.38 & 57514.47993 \\
o3o29  & 2013JL64 & 23.26$\pm$0.12 & 7.03 &         0.69$\pm$0.03 & 42.36 & 41.39 & 0.40 & 57515.53610 \\
o3o34 & 2013JH64 & 22.70$\pm$0.04 & 5.60 &          0.70$\pm$0.01 & 51.42 & 50.55 & 0.56 & 57133.53409 \\
o5s32 & 2015RJ277 & 23.21$\pm$0.04 & 7.12 &        0.63$\pm$0.01 & 39.62 & 38.78 & 0.76 & 57696.25289 \\
o5s45 & 2015RG277 & 23.15$\pm$0.03 & 6.79 &        0.97$\pm$0.02 & 42.51 & 41.62 & 0.60 & 57687.33189 \\
o5s05$\ssymbol{2}$ & 2015RV245 & 23.21$\pm$0.04 & 10.10 & 0.61$\pm$0.04 & 20.67 & 19.81 & 1.42 & 57696.36785 \\ 
o5s06$\ssymbol{2}$ & 2015RW245                  & 22.90$\pm$0.03 & 8.53 & 0.74$\pm$0.02 & 26.51 & 25.57 & 0.74 & 57688.30839 \\ 
o5s16PD$\ssymbol{2}$ & 2004PB112             & 22.99$\pm$0.03 & 7.39 & 0.77$\pm$0.04 & 35.62 & 34.76 & 0.82 & 57695.35870 \\ 
o5t04$\ssymbol{2}$ & 2015RU245                   & 22.99$\pm$0.04 & 9.32 & 0.81$\pm$0.01 & 22.53 & 21.65 & 1.17 & 57690.36639 \\ 
  \enddata
\tablenotetext{}{${}^*$OSSOS r$^{\prime}$ and ${\rm H_r}$ mag are at discovery, at an earlier observing geometry than the Col-OSSOS color measurement.} 
\tablenotetext{}{$\ssymbol{2}$Object with previously-published colors in \citet{Pike:2017gf} and reprocessed here using the latest version of TRIPPy \citep{Fraser:2016bd}.}
\tablenotetext{}{$\Delta$: heliocentric range, r: distance at discovery, $\alpha$: phase angle.}  
\end{deluxetable}

\section{Red dynamically excited TNOs have lower orbital inclinations}
\label{sec:results}

\begin{figure}[h!]
\centering
\includegraphics[angle=0, width=0.45\linewidth, trim=0cm 0cm 0cm 0cm, clip]{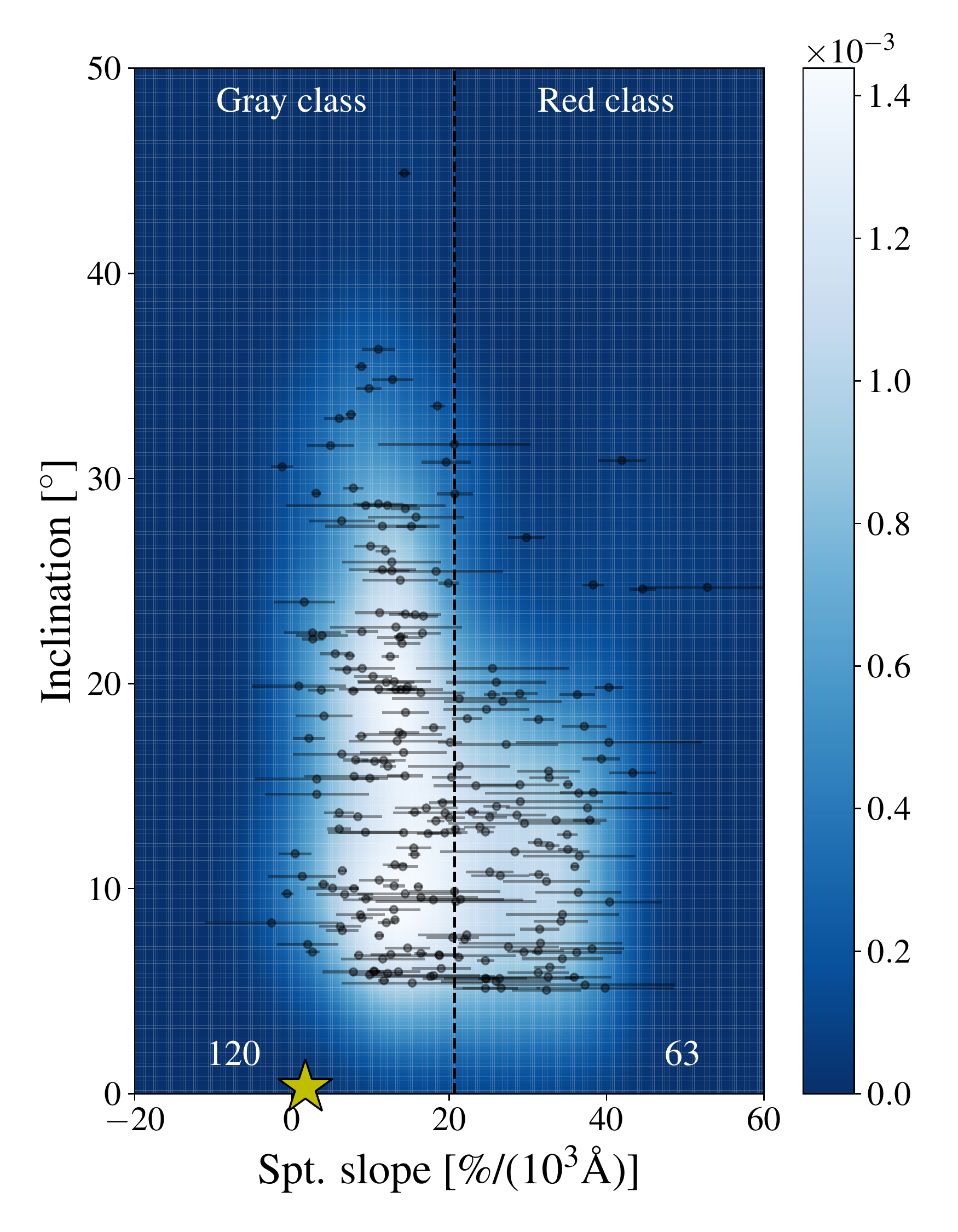}
 \caption{ Orbital inclination versus spectral slope of the dynamically excited TNOs and centaurs with i$>$5$\degr$. 
The vertical dashed line at ${\rm s=20.6\%/(10^3 \AA})$ marks the limit between the two TNO color classes as determined by the Gaussian Mixture Model. 
The yellow star indicates the spectral slope value of the Sun. 
The reddest object in our sample is Centaur (5145)~Pholus.
The number of classified gray and red objects is indicated at the bottom of the corresponding panel. 
Unclassified objects are those for which the error bar on the color measurement intersects with the division between the two color classes. 
A smoothed density plot is used to highlight the density of data points. 
Note the sparseness of red TNOs on highly inclined orbits. 
At 5$<$i$<$21$\degr$, where both color classes are well sampled, inclinations and spectral slopes are not correlated; 
a correlation would be expected in the context of a collision origin for TNO colors (section~\ref{sec:discussion}). 
Rather, TNOs appear to form two intrinsically different populations, each with its own orbital inclination and color distributions. }
\label{fig:color_inc_all}
\end{figure}

The whole dataset is shown in Figure~\ref{fig:color_inc_all}. 
The data suggest a relationship between color and inclination. 
In particular, there is a paucity of red TNOs with high orbital inclinations in our dataset.  
In the following sections, we assess whether this trend is statistically robust. 
To do so, we divide our dataset into two groups by color (section~\ref{sec:color_classes}) and test whether the two color groups have different inclination distributions (section~\ref{sec:tests1}). 
An alternative approach would be to divide our dataset into two inclination groups and comparing the color distribution of these two groups. 
We take the first approach as our dataset is clearly bimodal in color, as discussed in the next section. 
Effects of survey biases on our analysis are considered in section~\ref{sec:biases}. 

\subsection{Color classification}
\label{sec:color_classes}

\begin{figure}[h!]
\centering
\includegraphics[angle=0, width=0.45\linewidth, trim=0cm 0cm 0cm 0cm, clip]{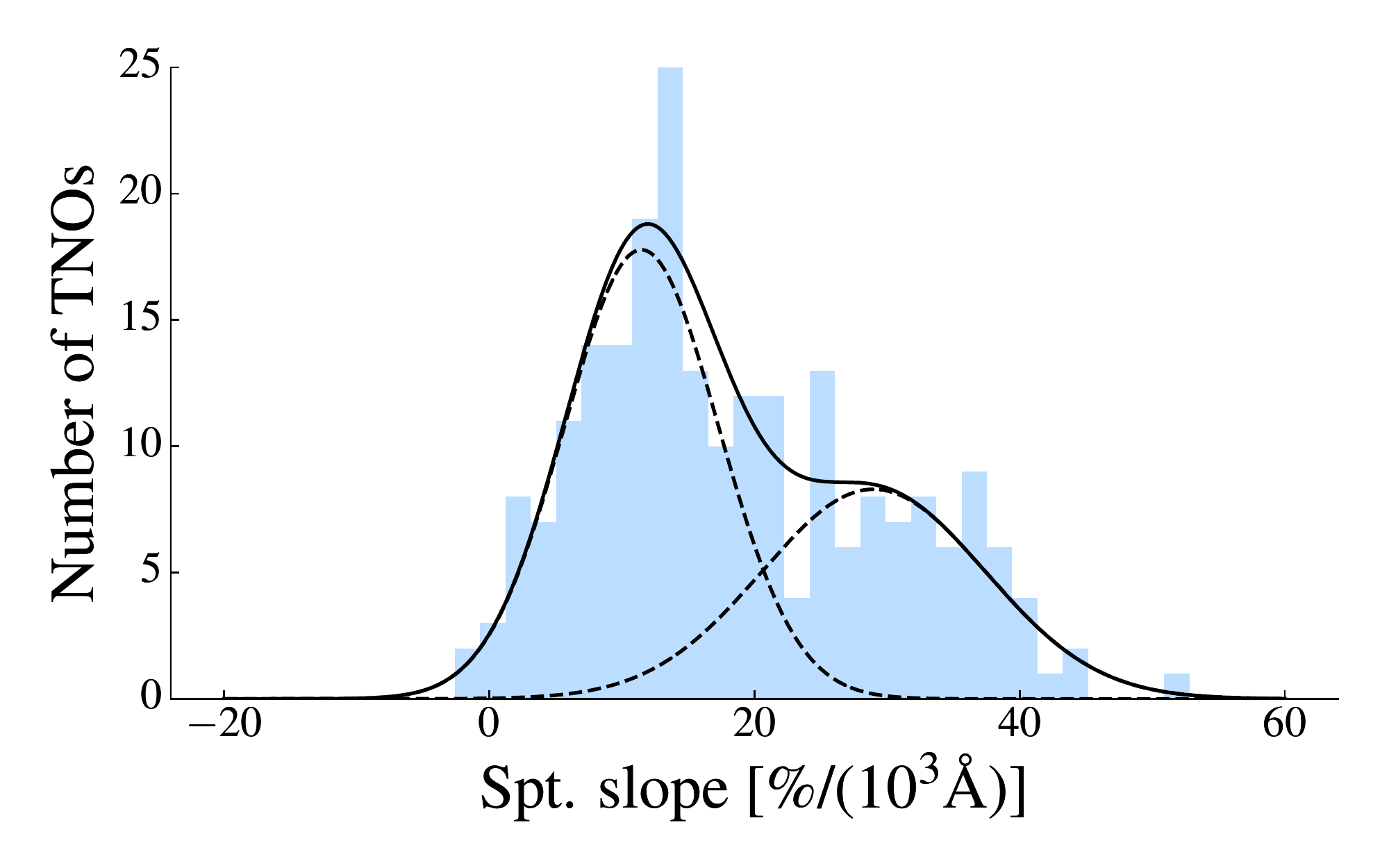}
 \caption{Histogram of spectral slope (s) for our dataset (229 objects in total). Application of a Gaussian Mixture Model decomposes our dataset into two normal distributions (dashed lines) that intersect at ${\rm s = 20.6\%/(10^3 \AA})$. We adopt this value to classify our dataset into the gray and red color classes. Table~\ref{tab:slope} shows our value for this division is consistent with those of previous works.}
\label{fig:slope_hist}
\end{figure}

Visually, the overall color distribution of all dynamically excited TNOs in our sample appears to be bimodal (Figure~\ref{fig:slope_hist}). 
The test of normality rejects a single gaussian distribution at the 99.8\% confidence level. 
We therefore test if it can be decomposed into multiple normal distributions. 
We calculated the Bayesian information criterion (BIC) for a set of N-component Gaussian Mixture Models (GMM), and confirmed that the best BIC is found for a two-component model.
The GMM decomposes our sample into two normal distributions, with respective means ${\rm s=11.5\%/(10^3 \AA})$ and ${\rm s=29.0\%/(10^3 \AA})$, and standard deviations ${\rm \sigma_s=5.8\%/(10^3 \AA})$ and ${\rm \sigma_s=8.4\%/(10^3 \AA})$. 
This confirms a previously known property of the Kuiper belt, which is that the dynamically-excited red TNOs exhibit a broader range of colors than the gray ones. 
The origin of this feature remains unclear (e.g., \citealt{Brown:2011}), and we do not investigate it further, as this goes beyond the scope of this paper.
The two normal distributions intersect at ${\rm s=20.6\%/(10^3 \AA})$, corresponding to optical colors (\emph{g}-\emph{r})\,=\,0.78, (V-R)\,=\,0.56 and (F606w-F814w)\,=\,-0.11. 
Therefore, we classify all objects with ${\rm s+\delta\,s<20.6\%/(10^3 \AA})$ as gray and all objects with ${\rm s-\delta\,s>20.6\%/(10^3 \AA})$ as red, where ${\rm \delta\,s}$ is the uncertainty on the spectral slope. 
Objects at the boundary between the two color groups were not classified as they are ambiguous. 
Our classification is consistent with previous works that placed the bifurcation between gray and red TNOs at comparable spectral slope values (Table~\ref{tab:slope}). 

\begin{deluxetable}{ccl}[h!]
\tabletypesize{ \scriptsize}
\tablecaption{ Adopted values for the bifurcation between the two color classes of dynamically excited TNOs. }
\tablehead{ \multicolumn{1}{c}{ Color } & \multicolumn{1}{c}{ Spectral slope } & \multicolumn{1}{l}{ Ref. }  \\
\multicolumn{1}{c}{  } & \multicolumn{1}{c}{ (${\rm \%/(10^3 \AA})$) } & \multicolumn{1}{l}{ }  }
\startdata
F606w-F814w=-0.15 & 18.0 &  \citealt{Fraser:2012cs}  \\ 
B-R=1.60 & 23.8 &  \citealt{Peixinho:2012bt}  \\ 
B-R=1.45 & 17.4 &  \citealt{Tegler:2016ep}  \\ 
\emph{g}-\emph{i}=1.17 & 22.8 &  \citealt{Wong:2017ck}  \\ 
\emph{g}-\emph{r}=0.78 & 20.6 & This work \\ 
\enddata
\tablenotetext{}{ }
\label{tab:slope}
\end{deluxetable}

\subsection{Statistical discrepancy between the inclination distributions of gray and red TNOs}
\label{sec:tests1}

\subsubsection{Statistical tests}

Four main tests were used to assess the orbital inclination versus spectral slope distribution of our dataset. 
The one-dimensional two-sample (2-s) Kolmogorov-Smirnov (KS; \citealt{Chakravarti:1967vt}) and Anderson-Darling (AD; \citealt{Stephens:1974db}) tests were used to determine whether the inclination distributions of the two TNO color groups defined in section~\ref{sec:color_classes} come from a single parent distribution. 
These two tests are broadly similar and also complementary. 
The KS test is more sensitive to the center of the distributions, while the AD test gives more weight to the tails. 
The AD test is generally considered more powerful than the KS test.  
These two tests are nonparametric with no reference to a Gaussian. 
We therefore report 1 minus the p-value of the test (in percentage) as the level of confidence that the two distributions are different, rather than a number of $\sigma$ deviation from a mean. 
We also used the Student's T-test of means \citep{Snedecor:1989tc} to determine whether the mean inclinations of the gray and red TNOs are equal. 
Finally, we derived the Spearman coefficient \citep{Zwillinger:2000} of our dataset to search for a correlation between orbital inclination and spectral slope. 
Unlike the other three tests, the Spearman test does not require dividing our dataset into two color groups. 
All results discussed in the following sections are summarised in Table~\ref{tab:stats1}. 
p-values were determined by computing each statistic for sets of inclination/spectral slope pairs constructed by independently randomly sampling the observed distribution of each parameter \citep{EfrTib93}. 
We sampled with replacement so the same value of inclination or spectral slope could be sampled several times. 
The reported confidence level (CL) then indicates the probability that a random sample produces a larger test statistic than the observed population. 

\subsubsection{Full dataset}
\label{sec:full}

We first consider the full dataset as a whole. 
All statistical tests indicate a strong (CL$>$99.7\%) discrepancy between the inclination distributions of the gray and red TNOs, as well as between their mean inclinations. 
The Spearman test indicates that spectral slope anti-correlates with inclination at the same significance level (Table~\ref{tab:stats1}).
The color versus inclination trend is particularly obvious when comparing the ratios of gray--to--red objects between low and high-inclinations. 
From Figure~\ref{fig:color_inc_all}, it appears that the inclination transition from where gray and red TNOs are roughly equally numbered to where red TNOs are almost nonexistent is located around i$=$21$\degr$. 
We therefore chose to compare the ratios of red--to--gray objects above and below this value. 
Only five TNOs -- i.e., 11\% of the objects -- can be unambiguously classified as red at i$>$21$\degr$ (Figure~\ref{fig:color_inc_all}). 
In contrast, red objects account for 42\% of the low-inclination (5$<$i$<$21$\degr$) population. 
To assess the likelihood of producing such an observation from a random population of objects, we bootstrap with replacement from the observed TNO samples of $N_g$ and $N_r$ objects, where $N_g$ and $N_r$ are the observed number of gray and red TNOs. 
We find that less than 0.1\% of the simulations produce a higher or equal ratio of gray--to--red objects than in the observed sample of high-i (i$>$21$\degr$) TNOs. 

While the above tests confirm a trend in color versus inclination, they cannot determine whether this trend results from the two color classes having different inclination distributions or from a direct correlation between color and inclination. 
To assess the possibility of a correlation, we consider only objects with 5$<$i$<$21$\degr$, where both color classes are thoroughly sampled by surveys (Figure~\ref{fig:color_inc_all}). 
Using the Spearman test, we find no evidence for a correlation between color and inclination for these objects (Table~\ref{tab:stats1}). 
This indicates that the Spearman test finds a correlation for all (low-i and high-i) TNOs because the number of red TNOs drops off near 21$\degr$ inclination. 
It follows that spectral slope and inclination of the dynamically excited TNOs are not directly correlated. 
Rather, red and gray objects have different inclination distributions. 
This opens interesting prospects for our understanding of the origin of TNO colors (see further discussion in section~\ref{sec:discussion}). 

\subsubsection{Individual populations}

\begin{figure}[h!]
\centering
   \begin{minipage}[t]{\linewidth}
	\centering
	\includegraphics[angle=0, width=0.9\linewidth, trim=0cm 0cm 0cm 2cm, clip]{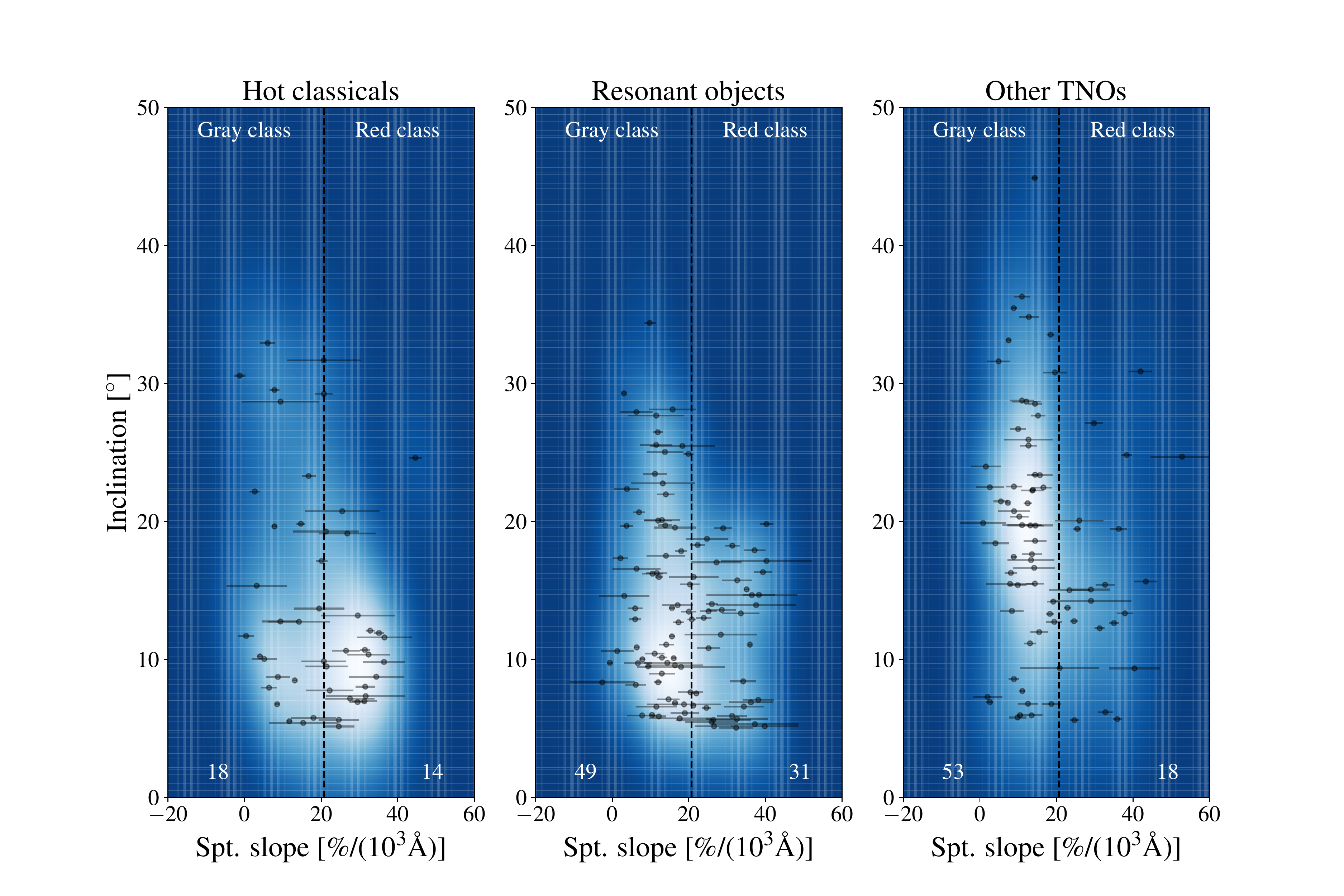}
   \end{minipage}
   \begin{minipage}[t]{\linewidth}
	\centering
  	\vspace{-0mm}
	\includegraphics[angle=0, width=0.9\linewidth, trim=0cm 0cm 0cm 2cm, clip]{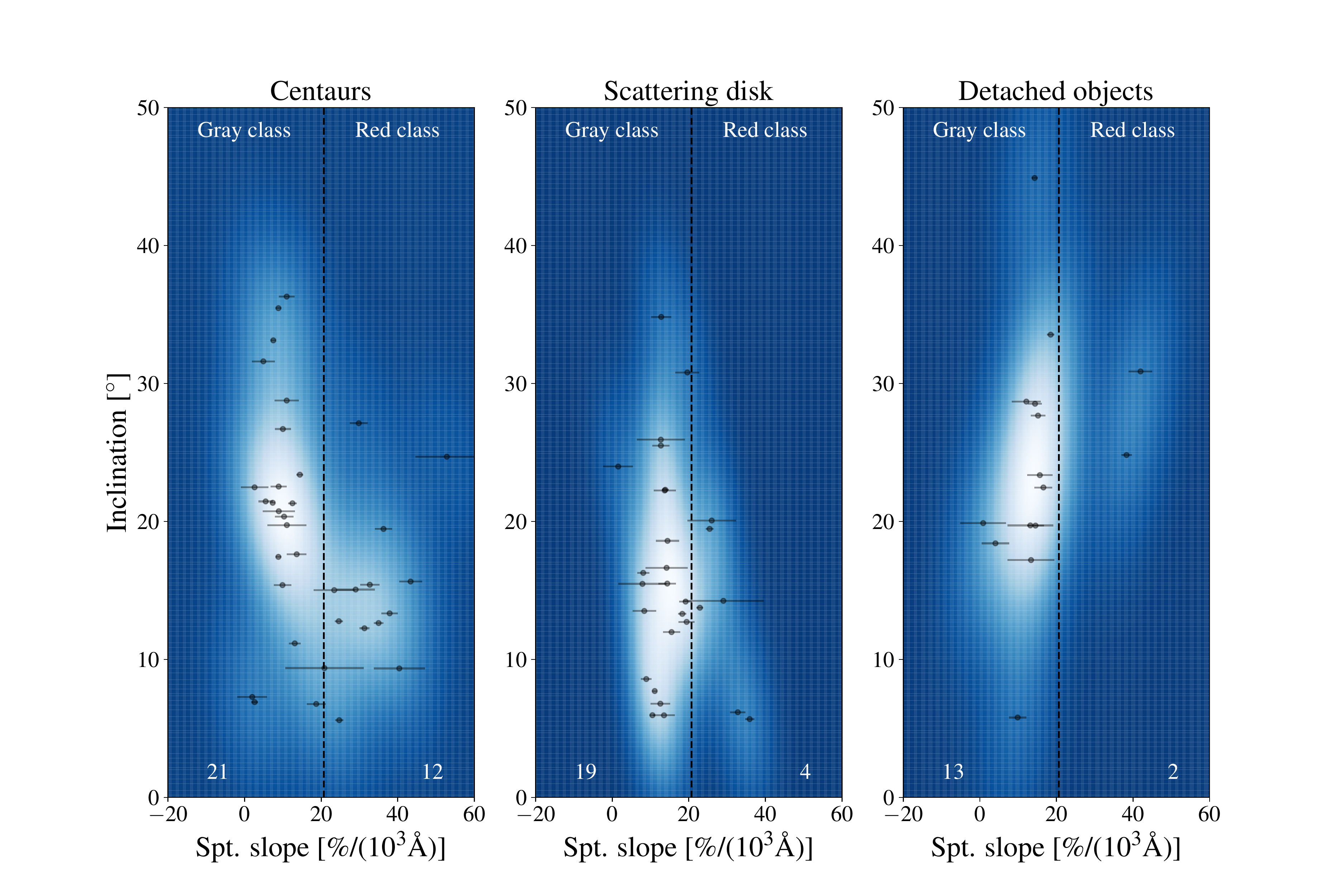}
	  \caption{ Same as Figure~\ref{fig:color_inc_all} for the individual dynamical populations of the dynamically excited TNOs. 
	  {\bf Top:} {\it left:} Hot classicals, {\it middle:} Resonant objects, {\it right:} Combined group of centaurs, scattering disk and detached objects. 
	  {\bf Bottom:} The combined group is divided into {\it left:} centaurs, {\it middle:} scattering disk objects, {\it right:} detached objects. 
The absence of red TNOs on highly inclined orbits appears to be a common property to all dynamical populations. 
Centaurs are the only population for which different orbital inclinations of the two color classes have been shown so far \citep{Tegler:2016ep}. Our dataset reveals a similar trend to that of \citet{Tegler:2016ep} for hot classicals, resonant objects and centaurs, but with a lower statistical significance (see Table~\ref{tab:stats1}). 
The small number of scattering disk and detached objects in our sample does not permit us to statistically confirm the existence of a matching trend in those populations. However, we do observe a lack of red scattering disk objects on highly inclined orbits (i$>$21$\degr$). \label{fig:color_inc}}
   \end{minipage}
\end{figure}

Here, we test whether the observed inclination distribution discrepancy  between gray and red TNOs holds for the individual dynamical subpopulations of objects. 
We consider three distinct groups of dynamically excited TNOs. 
The first two groups, the hot classicals and the resonant objects, are each considered separately as they comprise enough objects to allow a statistically robust analysis. 
The remaining objects, including centaurs, scattering disk, and detached objects, are considered as a single group. 
We emphasise that considering those objects together does not mean we assume they have a common origin. 
We adopt this approach because these populations, especially the scattering disk and detached objects, are too sparsely sampled in our dataset to be analyzed individually. 
Our goal here is to check whether the inclination discrepancy between gray and red TNOs holds for the remaining TNOs in our dataset. 
Nevertheless, it is believed that the scattering disk directly feeds the centaur population \citep{Duncan:1997hg}. 
Since objects from the Kuiper belt roughly preserve their inclinations as they become centaurs \citep{Volk:2013gv}, this seems the most sensible class merger. 
Figure~\ref{fig:color_inc} shows the inclination versus color distribution for the individual populations of TNOs. 

Applying previous statistical tests to the individual groups, we find that the discrepancy between the mean inclinations of the gray and red TNOs holds with CL=96.1\% on average for the hot classicals (Table~\ref{tab:stats1}). 
The Spearman correlation between inclination and spectral slope is less significant (CL=93.2\%). 
The KS and AD test return somewhat weaker results but this slightly depends on the magnitude and inclination cuts used to define our sample (see appendix~\ref{sec:stats}). 
Our results therefore are in agreement with previous works that reported an inclination versus color correlation for classical TNOs \citep{Tegler:2000fx, Trujillo:2002ew, Hainaut:2002fx, Doressoundiram:2005ha, Peixinho:2008gn}. 
These works however could have been contaminated by the presence in the sample of objects belonging to peculiar compositional groups, including cold classical objects. 
Here, we show that the observed discrepancy between gray and red hot classicals holds when interlopers are removed from this population. 

A similar trend is found in the resonant group and in the group of centaurs/scattering/detached objects at a comparable significance level (Table~\ref{tab:stats1}). 
We also observe that out of the 102 resonant TNOs in our dataset -- where 49 are gray and  31 are red (the rest being unclassified) -- the 14 highest-inclination objects all belong to the gray class. 
When bootstrapping two random samples of 49 and 31 objects from the resonant population, we find that this occurs in less than 0.1\% of the simulations, thus strongly supporting the conclusion that the gray and red resonant TNOs have different inclination distributions. 
Finding that resonant objects follow the same trend as other dynamical populations of TNOs is interesting. 
It implies that the initial orbital inclinations these objects had before being captured into orbital resonances with Neptune were not completely randomized by the capture process and the dynamical mechanisms, such as Kozai oscillation (e.g., \citealt{Gomes:2005fl}), that take place in those resonances. 

We stress that the group of centaurs/scattering/detached TNOs is strongly biased by the presence of centaurs in the sample. 
Excluding those objects completely removes the trend of gray and red TNOs having different orbital inclinations in that group. 
We note however that the scattering disk population does not contain any red objects at i$>$21$\degr$, similar to what is observed in the resonant population and, to a lesser extent, the hot classical (1 interloper) and the centaur (2 interlopers) populations (Figure~\ref{fig:color_inc}). 
Scattering disk objects might therefore follow the same trend as other TNO populations, but the current small number of scattering TNOs with well-measured colors prevents us from drawing any firm conclusion. 
Detached objects seem to constitute the only population that does not follow the trend, but their sample is also too small for conclusive analysis. 

\subsubsection{Summary of statistical tests}

To summarise, we divided our dataset into two color classes at ${\rm20.6\%/(10^3 \AA})$ and found a highly significant difference between the orbital inclination distributions of the gray and red TNOs. 
This discrepancy cannot be detected at the CL$>$99.7\% significance level in the individual dynamical population of TNOs, but  
it does show up in each of these populations at lower statistical significance. 
This indicates that 1) this trend is common to every population of dynamically excited TNOs and, 2) our overall dataset is not biased by one particular dynamical population. 
We emphasise that the results presented here slightly vary depending on the adopted magnitude and inclination cuts used to define the analyzed sample, as well as the spectral slope value of ${\rm20.6\%/(10^3 \AA})$ we adopt to differentiate between gray and red TNOs. We find however that a reasonable modification of those values does not affect the conclusions of our study in any significant way (see appendix~\ref{sec:stats}). 

\begin{deluxetable}{lccccc}[h!]
\tabletypesize{ \scriptsize}
\tablecaption{ Confidence level for gray and red TNOs having different inclination distributions.  }
\tablehead{\colhead{  } & \multicolumn{1}{c}{ hc } & \multicolumn{1}{c}{ res } & \multicolumn{1}{c}{ cen+sca+det } & \multicolumn{1}{c}{ all } & \multicolumn{1}{c}{ all--cen }  }
\startdata
2-sample KS-test                                                      &  98.1 \% &  91.2  \% &  97.8 \% &  {\bf $>$99.9} \% &  {\bf $>$99.9} \% \\ 
2-sample AD-test                                                      &  93.6 \% &  98.5  \% &  96.7 \% &  {\bf $>$99.9} \% &  {\bf $>$99.9} \% \\ 
T-test of means                                                         &  99.5 \% &  99.2  \% &  97.9 \% &  {\bf $>$99.9} \% &  {\bf $>$99.9} \% \\ 
Spearman correlation (i$>$5$\degr$)                       &  93.2 \% &  95.0  \% &  85.4 \% &  {\bf 99.7} \%       &  99.6 \% \\ 
Spearman correlation (5$\degr$$<$i$<$21$\degr$) &  60.3 \% &  65.5  \% &  82.2 \% &  82.1 \%              &  82.5 \% \\ 
\enddata
\tablenotetext{}{ hc=hot classicals, res=resonant objects, cen=centaurs, sca=scattering disk objects, det=detached objects, all=full dataset, all--cen=full dataset without centaurs. Bold indicates a $\geq$99.7\% confidence level. } 
\label{tab:stats1}
\end{deluxetable}

\subsection{Is the observed color-inclination distribution an artifact of the discovery surveys?}
\label{sec:biases}

\subsubsection{Analytical argument}
\label{sec:analytical}

The dataset built for this work uses photometric measurements from a large variety of discovery surveys, each with its own detection biases and discovery efficiency. This raises the question as to whether these observing biases could be responsible for the observed inclination discrepancy between gray and red TNOs reported in this work. 
A discovery bias in favor of red TNOs comes from the combination of their higher optical albedos \citep{Fraser:2014kp, Lacerda:2014hw} and redder colors with respect to gray objects. 
The observed ratio of red-to-gray objects depends on the filter used for the observations. 
Observations performed in broad-band filters at longer wavelengths will discover a higher fraction of red TNOs than observations made at shorter wavelengths. 
This can introduce a color-inclination correlation if low latitude discovery surveys (which are biased toward lower inclination TNOs) were performed at longer wavelengths than high latitude surveys. 
To test whether this effect can be responsible for the observed distribution, let's consider the worst-case scenario, which would be if off-ecliptic TNO surveys were made in V-band (550\,nm), while on-ecliptic surveys were made in R-band (660\,nm).

The number of objects with albedos between $a$ and $a+da$, between radii $R$ and $R+dR$, and heliocentric distances between $r$ and $r+dr$ detected by the survey is:

\begin{equation} 
n(R,r,a) = A\, f(R)\, g(r)\, h(a)\, da\, dr\, dR,
\end{equation}

\noindent where $f(R)$ is the TNO size distribution, g(r) is the Kuiper belt radial distribution, h(a) is the TNO albedo distribution, and $A$ is a constant. 
Assuming a power-law distribution in R, $f(R)\propto R^{-q}$, and recasting the expression for $R$ in terms of apparent magnitude $m$, it can be shown \citep{Schwamb:2018} that the number of objects with albedo $a_0 \leq a \leq a_1$ and magnitude $m_0 \leq m \leq m_1$ observed within the boundaries $r_0$ and $r_1$ of the Kuiper belt is given by:

\begin{equation}
N (m_0<m<m_1) = C \int_{a_o}^{a_1} h(a) a^{\frac{5\alpha}{2}} da \int_{r_o}^{r_1}  g(r) r^{-5\alpha} \Delta^{-5\alpha} dr 10^{\alpha m},
\end{equation}

\noindent where $C$ is a constant, $\Delta$ is the geocentric distance to the object, and $\alpha = (q-1)/5 \approx1.0$ \citep{Fraser:2014kp} is the power-law slope of the apparent magnitude distribution. 
Here, we have assumed that the albedo, radial, and size distributions are independent. 
For simplicity, this expression does not reflect the apparent latitudinal and longitudinal variation of TNO density of the resonant populations. 
We are only interested in the ratio of gray to red objects observed within a given survey. 
Assuming that within each orbital class, separate color populations share the same spatial distribution, the latitudinal and longitudinal variations will be reflected in both color populations equally. Thus, for our derivation such variations can be ignored. 
In the case of a population composed of two color groups, where each group has its own albedo distribution, but similar size and radial distributions, and where the intrinsic number of objects in the two groups is given by $A_2 = \gamma A_1$, where $\gamma$ is a constant, the observed ratio of populations 1 and 2 is:

\begin{equation}
\frac{N_1}{N_2}(m_0<m<m_1) = \gamma \frac{\int_{a_o}^{a_1} h_1(a) a^{\frac{5\alpha}{2}} da}{\int_{a_o}^{a_1} h_2(a) a^{\frac{5\alpha}{2}} da}.
\label{eq:eq3}
\end{equation}

Here, we have assumed that the fraction of objects in the two groups is not a function of the discovery survey's observing depth. 
This is true as long as the size distribution can be modelled as a single power law ($f(R)\propto R^{-q}$), i.e., as long as the limiting magnitude of the surveys is not sensitive to the break (or divot) magnitude (see, e.g., \citealt{Shankman:2013fx, Shankman:2016hi, Fraser:2014kp, Lawler:2018gf}). 
A fraction of objects in our dataset, mainly centaurs and scattering disk objects, are fainter than the magnitude break measured for these objects (H=7.7 or 8.3; \citealt{Lawler:2018gf}). 
The presence of a break in the size distribution is explored numerically below in section~\ref{sec:surveysim}. 


For simplicity, we consider the simple scenario where the two color groups each have a unique albedo, $a_1$ and $a_2$. 
Then, the albedo distribution is given by the Dirac delta function $h_n(a) = \delta(a_n - a)$, and equation\,\ref{eq:eq3} becomes:

\begin{equation}
\frac{N_1}{N_2}(m_0<m<m_1) = \gamma\,a_1^{\frac{5\alpha}{2}} a_2^{-\frac{5\alpha}{2}}.
\label{eq:eq4}
\end{equation}

In V-band (550\,nm), gray and red TNOs have mean albedos of 6\% and 12\%, respectively \citep{Fraser:2014kp, Lacerda:2014hw}, implying a red-to-gray ratio $\frac{N_{red}}{N_{gray}} = 5.66 \gamma$. 
In this work, we derived mean spectral slopes of 11.4\%/(10$^3$\AA) and 29.1\%/(10$^3$\AA) for the gray and red color classes of TNOs, respectively. 
This implies albedos of 6.7\% and 15.7\% in R-band (660\,nm), and $\frac{N_{red}}{N_{gray}} = 8.41\gamma$. 
Therefore, in the V spectral band the observed red--to--gray ratio is 67\% that in the R spectral band. 

In our dataset, we measure a red--to--gray ratio of 0.821 for 5$<$i$<$10\,$\degr$ (23 red objects, 28 gray, and 17 unclassified). 
Assuming the worst case scenario, where all on-ecliptic discovery surveys were performed in R and all off-ecliptic surveys were performed in V, the expected red--to--gray ratio at high inclination would be 0.550. 
Yet, the observed ratio for i$>$21\,$\degr$ is 0.119 (5 red objects, 42 gray, and 6 unclassified). 
If considering \emph{g} (480\,nm) and \emph{r}-band (620\,nm) filters instead of V and R, then the observed ratio in \emph{g} is 56\% that in \emph{r}, and the expected ratio in our dataset would be 0.450 at i$>$21\,$\degr$. 
Choosing different inclination cuts for the two subsamples slightly varies these numbers but does not affect the conclusion. 
Therefore, it appears that even the observationally heavily biased scenario considered here can not account for the observed relationship between color and inclination.

\subsubsection{Survey simulation}
\label{sec:surveysim}

We further test the effects of discovery biases on the color versus inclination distribution of TNOs 
using the OSSOS survey simulator \citep{Bannister:2016cp, Bannister:2018ha, Lawler:2018uy}. 
This simulator replicates the detection characteristics (pointings, detection and tracking efficiencies) of a given survey to determine which input model objects would be detected by the survey. 
We use this simulator to model a heavily biased survey composed of a single on-ecliptic block performed in r-band and two high-latitude blocks, with mean inclinations of 12$\degr$ and 22$\degr$, performed in g. 
All blocks are equally sensitive to the input objects in our simulation; 
Considering TNOs' mean (\emph{g}-\emph{r})=0.78 color, observational depth was set to m$_r$=24.0 for the on-ecliptic block, and m$_g$=24.75 for the off-ecliptic ones. 
We use as input populations the Canadian-France Ecliptic Plane Survey (CFEPS) L7 model of the Kuiper belt \citep{Petit:2011hj} and \citet{Shankman:2016hi}'s improved version of the \citet{Kaib:2011ky}'s (hereafter K11) model of the scattering disk and centaur populations. 

Each input object was cloned several times, allowing its orbital parameters $a$, $e$ and $i$ to vary by 10\% with respect to their nominal values, while randomising its position angles ($\omega$, $\Omega$ and M). 
This cloning ensures a sufficient number of simulated TNOs are randomly produced; we ran the simulator until a sample of 10,000 objects sampling the full parameter space was detected. 
Simulations were run independently for the on-ecliptic block and the high-latitude ones to ensure the same number of objects (5,000) were detected by the blocks. 
Optical (\emph{g}-\emph{r}) colors were drawn from the double normal distribution measured in section~\ref{sec:color_classes} and randomly attributed to the simulated objects.  
The input color ratio is irrelevant here; 
we are interested only in the {\it relative} color ratios between low-i and high-i objects in the output survey population. 
In order to determine the detectability of the simulated objects, an absolute magnitude distribution was assumed in the magnitude range of interest (5$<$${\rm H_r}$$<$11).  
We used two different distributions: a single slope distribution with exponent $\alpha$=1.0, and a broken-slope distribution, with slope exponents $\alpha_b$=1.0 and $\alpha_f$=0.2 for the bright and faint end of the distribution, respectively \citep{Fraser:2014kp}. 
The break magnitude of the broken-slope distribution was set to ${\rm H_{r}}$=7.7 for the gray TNOs and ${\rm H_{r}}$=6.9 for the red ones, in order to account for their respective mean albedos of 6\% and 12\%. 

\begin{figure}[h!]
\centering
\includegraphics[angle=0, width=0.38\linewidth, trim=0cm 0cm 0cm 0cm, clip]{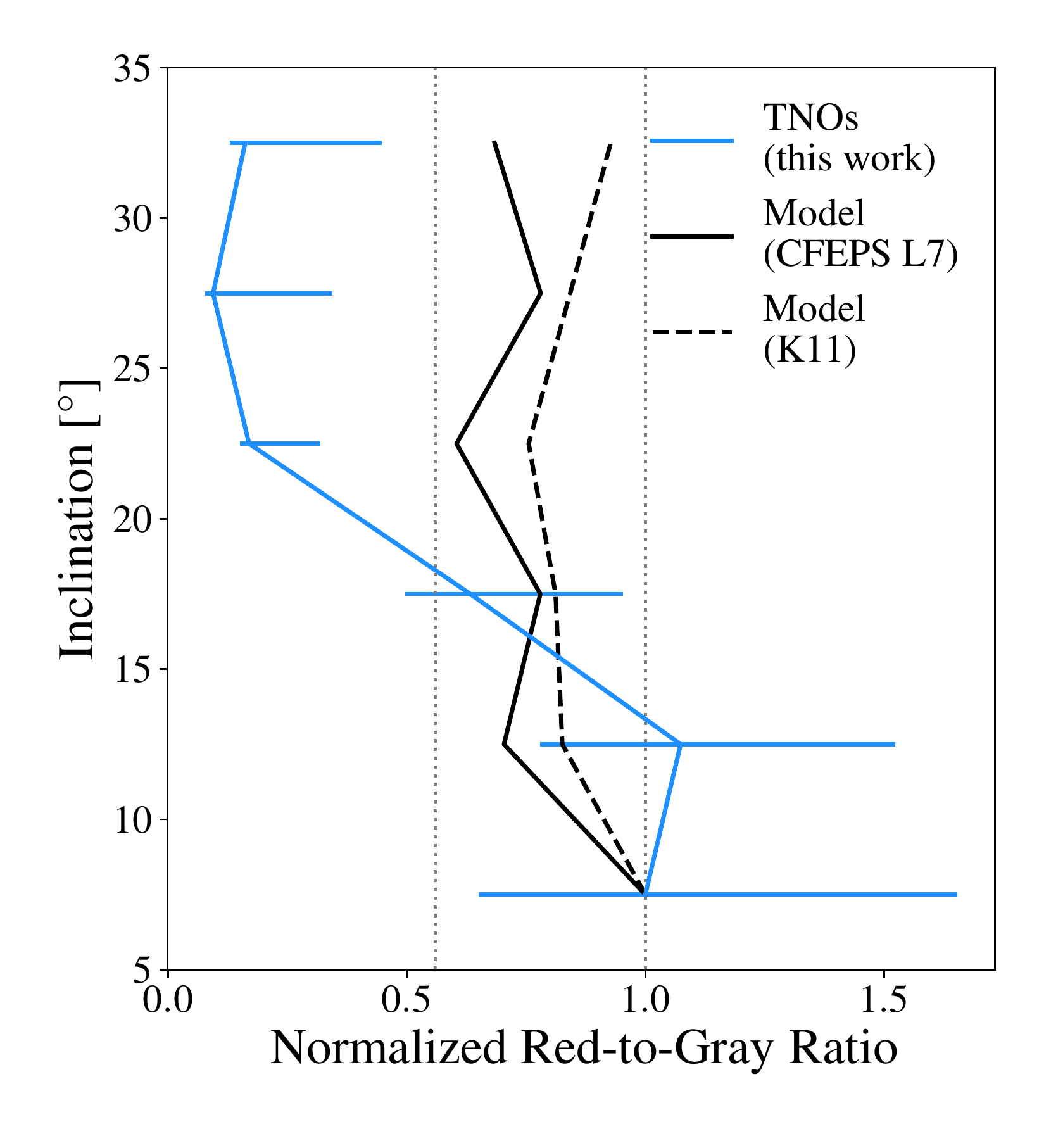}
 \caption{ 
The inclination-dependent color ratio observed in our TNO sample (blue) compared to models of a heavily color-biased survey (section~\ref{sec:surveysim}; black). 
The survey model cannot reproduce the observed color ratio, implying it is intrinsic to the TNO population.
Error bars on the TNO sample are derived from the number of color-unclassified TNOs (section~\ref{sec:color_classes}) in each inclination bin of 5$\degr$. The simulated survey observed a uniform color-ratio model on the ecliptic in \emph{r} and at high latitudes in \emph{g}. The indents in model color ratios at i$\approx$12, 22$\degr$ correspond to the mean inclinations of our simulated survey high-latitude sky blocks. Dotted vertical lines indicate theoretical extreme color-ratio cases observable for our simulation (section~\ref{sec:analytical}).
 }
\label{fig:simulation1}
\end{figure}

The population of detected objects in our simulated survey reveals a trend of increasing color with decreasing inclination (Figure~\ref{fig:simulation1}), in agreement with our survey setup. 
We first consider the population drawn from the single-slope distribution. 
The red-to-gray ratio of the objects found in the on-ecliptic block is 54\% that in the off-ecliptic blocks for the L7 model, and 62\% for the K11 model, in good agreement with the predicted value of 56\% derived in section~\ref{sec:analytical}. 
However, a fraction of the high-i TNOs in our simulation are detected by the low-latitude block. 
Because of this, the color discrepancy between the low-i and high-i objects in our simulation is weaker than predicted analytically (section~\ref{sec:analytical}); 
The red-to-gray ratio of the high-inclination ($>$21\,$\degr$) objects is 0.689 that of the low-inclination (i$<$10\,$\degr$) ones for the L7 model, and 0.854 for the K11 model. 
This strengthens our proposal that survey biases can not account for the observed color versus inclination distribution. 
Finally, we find that the presence of a break in the magnitude distribution increases the red-to-gray ratio of the population detected in the high-latitude blocks relative to the on-ecliptic one; 
the on-ecliptic block ratio is 63\% that in the off-ecliptic blocks for the L7 model, and 69\% for the K11 model. 
This reinforces our conclusion that the observed color versus inclination correlation cannot be a result of the survey detection biases.

\subsubsection{Real-life surveys}
\label{sec:isthisreallife}

We assess a subset of past surveys to further show that the correlation we report of color versus inclination distribution of the dynamically-excited TNO populations is not the result of a discovery bias. 
Indeed, real surveys are far from being as biased as in the worst-case scenario we explored in Sections~\ref{sec:analytical} and \ref{sec:surveysim}. 
 We select the surveys with the most discoveries and an obtainable pointing history, including \citet{Chiang1999, Gladman1998, Gladman2001, Trujillo2001, Allen2002, Bernstein:2004kn, Petit2006, Fuentes2008, Fuentes2009, Fraser2009, Fraser2008, Fraser2010, Alexandersen:2016}, the Deep Ecliptic Survey (DES; \citealt{Millis2002, Elliot2005}), CFEPS \citep{Petit:2011hj, Petit2017}, and OSSOS \citep{Bannister:2016cp, Bannister:2018ha}. 
This list is by no means exhaustive, but we ensure it covers well the range of ecliptic latitudes and of filters sampled by past surveys, and it includes the surveys that have reported the most detections of TNOs with H$_r$$>$5.0 (the DES, CFEPS and OSSOS). 
We omit the two surveys of \citet{Schwamb:2010} and \citet{Rabinowitz:2012}, which were up to large ecliptic latitudes, in red-band and in broad-band red+blue filters respectively, as we could not retrieve their precise pointing history. 

\begin{figure}[h!]
\centering
\includegraphics[angle=0, width=0.9\linewidth, trim=0cm 0cm 0cm 0cm, clip]{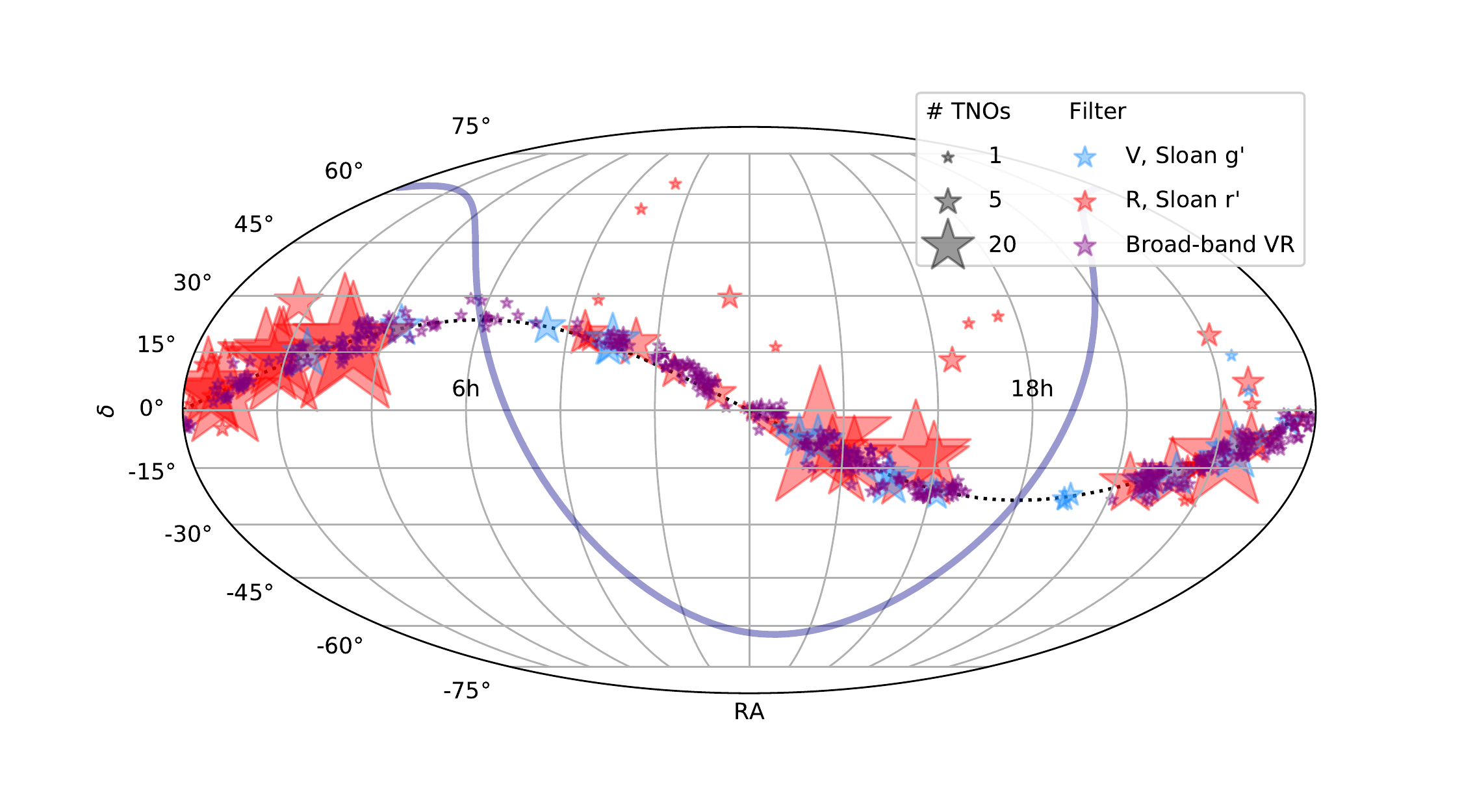}
 \caption{On-sky pointing coordinates of the TNO discovery surveys listed in Section~\ref{sec:isthisreallife}. 
We show the observing filter in which these surveys were performed, and schematically indicate their number of discoveries. 
Only the survey fields with positive detections are displayed for clarity. 
The dotted black curve and the blue curve indicate the location of the ecliptic and the galactic planes, respectively. 
Almost all surveys performed with blue-band and broad-band blue+red filters were conducted close to the ecliptic plane. 
Most high-latitude observations come from the High-Latitude Component of CFEPS \citep{Petit2017}, which was performed with a red-band filter. 
 These past observations imply that trans-Neptunian objects detected on highly-inclined orbits should be preferentially red compared to the low-inclination ones. 
This is contrary to the correlation we report. 
 }
\label{fig:pointing}
\end{figure}

Figure~\ref{fig:pointing} highlights that these surveys did not favor the detection of red TNOs on highly-inclined orbits. 
On the contrary, most surveys performed far from the ecliptic plane used red-band filters, whereas the majority of surveys using blue-band and broad-band blue+red filters were conducted close to the ecliptic. 
We should therefore expect objects detected on highly-inclined orbits to be preferentially red compared to the low-inclination ones, which is contrary to the correlation we report. 
We also note that only a small fraction of the known TNOs were detected in the high-latitude fields, in agreement with their narrow orbital inclination distribution. 
This implies that even if a discovery bias exists in our dataset, it probably only has a marginal effect on its color-inclination distribution.

\section{Discussion}
\label{sec:discussion}

Considering the commonly accepted idea that centaurs are dynamically evolved TNOs, it would appear surprising if the inclination dichotomy between gray and red centaurs found by \citet{Tegler:2016ep} did not occur in the Kuiper belt. Indeed, how would such a signal appear after the scattering of centaurs from the Kuiper belt?
Here, we have shown that this dichotomy also exists in the Kuiper belt, and that it appears to be a common feature to all dynamical classes of the dynamically hot TNOs. This removes the conundrum of centaurs being the only population observed so far to exhibit that property. 

Finding different orbital properties for the gray and red dynamically hot TNOs opens up interesting prospects for understanding the origin of the color diversity of small TNOs. So far, two main interpretations have been put forward to explain this diversity. 
The first hypothesis suggests that all TNOs were originally similar, and evolutionary -- collisional and resurfacing -- processes altered them differently. 
The second hypothesis is that the two TNO color classes are composed of intrinsically different objects, with distinct compositions, that probably formed at different locations in the protoplanetary disk. 

Orbital inclination is directly linked to the strength of the collisional environment. 
As such, previous reports of a color versus inclination correlation in the classical belt \citep{Tegler:2000fx, Trujillo:2002ew, Hainaut:2002fx, Doressoundiram:2005ha, Peixinho:2008gn} were interpreted to be the result of collisional resurfacing processes \citep{GilHutton:2002bh, Trujillo:2002ew, Stern:2002bj, Moroz:2003kl, Delsanti:2004jg, SantosSanz:2009gq}. 
In that scenario, we would expect a direct correlation between color and inclination because the collisional kinetic energy directly correlates with inclination.
Yet, we find no such correlation in our dataset for orbital inclinations 5$<$i$<$21$\degr$, where both color classes are well sampled by surveys (section~\ref{sec:full}). 
As such, the strong Spearman correlation found for the complete set of TNOs most likely comes from the absence of red TNOs above 21$\degr$ inclination.
The lack of TNO color variation with rotation \citep{Jewitt:2001hu} and the lack of a correlation between color and orbital eccentricity \citep{Thebault:2003fj, Thebault:2003bc} also appear to contradict a collisional origin for the inclination versus color relationship. 
Numerical simulations reveal that the kinetic energy TNOs receive from collisions always correlates better with eccentricity than inclination. 
Applying the same statistical tests as presented in Section~\ref{sec:tests1} to the eccentricity versus color distribution in our dataset reveals no correlation between these parameters.

It therefore appears that the dynamically excited TNOs formed in two intrinsically different populations, each with its own orbital inclination and color distributions. 
Dynamical simulations have shown that the high-inclination TNOs could have formed closer to the Sun and been emplaced at their current location following a phase of planetary migration \citep{Gomes:2003cd}.
Our study therefore suggests that gray TNOs accreted closer to the sun compared to red objects. 
During migration, their orbits were more excited by Neptune, which widened their orbital inclination distribution. 
The presence of gray class interlopers in the cold classical region, the so-called `blue binaries' \citep{Fraser:2017kh}, may however complicate this picture. 
The fact that these interlopers are all gray suggests that the gray class originally bordered the inner edge of the cold classical region \citep{Schwamb:2018}. 
Future detailed modeling of planetary migration in a color-varying planetesimal disk will provide predictions which can be tested against our observed color-inclination distribution. 

Different formation locations and emplacement efficiency as the origin of the two TNO color groups also appears compatible with the observation that gray TNOs are proportionally more numerous in the dynamically excited TNO populations. 
In our dataset, gray TNOs represent 61\%, 64\%, 83\% and 87\% of the total number of objects in the resonant, centaur, scattering and detached populations, respectively, whereas they represent `only' 56\% of the hot classicals. 
We however stress that this observation must be considered with extreme caution as it relies on a heavily-biased dataset built from various surveys with different detection biases. 
The Col-OSSOS survey will, upon completion, provide a complete fully debiased sample of TNO colors from which the intrinsic ratio of gray--to--red objects in each dynamical population will be derived \citep{Schwamb:2018}. 

\section{Conclusion}

We report new (\emph{g}-\emph{r}) colors for 25 small (H$_{\rm mag}$$>$5) dynamically excited TNOs observed in the Col-OSSOS survey. 
Combined with previously published color measurements from Col-OSSOS and other surveys, we consider the colors of a set of 229 dynamically excited TNOs and centaurs. 
We show that, when dividing our dataset into two color classes at spectral slope ${\rm s=20.6\%/(10^3 \AA})$, red objects have a lower orbital inclination distribution than their gray counterparts. 
This is true whether or not centaurs are included in our sample. 
This trend appears to be common to every dynamical population in the Kuiper belt. 
The scattering and detached populations are too sparsely sampled to confirm a matching trend in those populations. 
Even in the worst-case scenario, observing biases in the discovery surveys cannot account for this correlation: it is intrinsic to the underlying TNO and centaur populations. 
Considering that TNOs are the precursors of centaurs, and that their inclinations are roughly preserved as they become centaurs \citep{Volk:2013gv}, our finding solves the conundrum of centaurs being the only outer Solar System population identified so far to exhibit this property \citep{Tegler:2016ep}. 
Finally, the observed inclination difference between gray and red TNOs tends to favor the hypothesis of different formation locations for these objects, and a compositional gradient in the original protodisk of planetesimals, rather than different evolutionary pathways.

\section*{Acknowledgements}

The authors acknowledge the sacred nature of Maunakea, and appreciate the opportunity to observe from the mountain. 
This work is based on observations from the Large and Long Program GN-LP-1 (2014B through 2016B), obtained at the Gemini Observatory, which is operated by the Association of Universities for Research in Astronomy, Inc., under a cooperative agreement with the NSF on behalf of the Gemini partnership: the National Science Foundation (United States), the National Research Council (Canada), CONICYT (Chile), Ministerio de Ciencia, Tecnolog\'{i}a e Innovaci\'{o}n Productiva (Argentina), and Minist\'{e}rio da Ci\^{e}ncia, Tecnologia e Inova\c{c}\~{a}o (Brazil). 
We thank the staff at Gemini North for their dedicated support of the Col-OSSOS program. 
Data was processed using the Gemini IRAF package. 
STSDAS and PyRAF are products of the Space Telescope Science Institute, which is operated by AURA for NASA.
This research used the facilities of the Canadian Astronomy Data Centre operated by the National Research Council of Canada with the support of the Canadian Space Agency. 
The authors thank C. Shankman for sharing his Python implementation of the OSSOS survey simulator, and S. Lawler and N. Kaib for providing their model of the scattering disk and centaur populations. 
M.T.B. appreciates support from UK STFC grant ST/L000709/1. 
M.E.S. was supported by Gemini Observatory. 
K.V. acknowledges support from NASA grants NNX14AG93G and NNX15AH59G. 
N.P. acknowledges funding from the Portuguese FCT -- Foundation for Science and Technology (ref: SFRH/BGCT/113686/2015). CITEUC is funded by National Funds through FCT -- Foundation for Science and Technology (project: UID/ Multi/00611/2013) and FEDER -- European Regional Development Fund through COMPETE 2020 -- Operational Programme Com- petitiveness and Internationalisation (project: POCI-01-0145-FEDER-006922).

\facilities{Gemini:Gillett (GMOS-N, NIRI)}
\software{Astropy \citep{Collaboration:2013cd}, IRAF \citep{Tody:1986df}, Matplotlib \citep{Hunter:2007}, NumPy \citep{vanderWalt:2011dp}, OSSOS survey simulator \citep{Lawler:2018uy}, pyraf, Scikit-learn \citep{scikit-learn}, SciPy \citep{vanderWalt:2011dp}, Sextractor \citep{Bertin:1996hf}, STSDAS, TRIPPy \citep{Fraser:2016bd}}

\bibliographystyle{apj}
\bibliography{references} 

\appendix

\section{Full dataset}
\label{sec:dataset}

\startlongtable
\begin{deluxetable}{llccccccl}
\tabletypesize{ \scriptsize}
\tablecaption{\label{tab:sptslopes}}
\tablehead{\colhead{MPC} & \colhead{Target} & \colhead{Orbit} & \colhead{H${_r}$} & \colhead{ a (au) } &\colhead{e} & \colhead{ inc ($\degr$) } &  \colhead{ Spectral slope } &  \colhead{ Ref.} \\
\colhead{number} & \colhead{Designation} & \colhead{Classification} & \colhead{} & \colhead{ } &\colhead{ } & \colhead{ } &  \colhead{ (${\rm \%/10^3 \AA}$) } &  \colhead{ } } 
\startdata
\multicolumn{9}{c}{ Hot Classical } \\
181708  &              1993FW &      cla (i) & 6.85 &  43.72 & 0.05 & 7.75 &  22.23$\pm$6.16 &  2 \\
  &            1996RQ20 &      cla & 7.17 &  43.86 & 0.10 & 31.67 &  20.64$\pm$9.66 &  2 \\
  &            1996TS66 &      cla & 6.23 &  43.89 & 0.13 & 7.35 &  31.62$\pm$10.39 &  2 \\
  &            1997CV29 &      cla (i) & 7.45 &  42.18 & 0.05 & 8.03 &  31.50$\pm$2.50 &  2 \\
  &             1997QH4 &      cla & 7.28 &  42.70 & 0.08 & 13.19 &  29.57$\pm$9.68 &  2 \\
385191  &             1997RT5 &      cla & 7.21 &  41.31 & 0.03 & 12.73 &  14.20$\pm$8.10 &  2 \\
  &           1998FS144 &      cla & 6.99 &  41.77 & 0.02 & 9.87 &  20.68$\pm$5.84 &  2 \\
181855  &            1998WT31 &      cla & 7.68 &  45.98 & 0.18 & 28.68 &  9.38$\pm$10.16 &  2 \\
  &           1999CB119 &      cla & 7.04 &  46.91 & 0.13 & 8.75 &  34.40$\pm$7.22 &  2 \\
  &           1999CL119 &      cla & 6.00 &  46.87 & 0.01 & 23.29 &  16.71$\pm$1.82 &  2,3 \\
59358  &           1999CL158 &      cla (i) & 6.88 &  41.44 & 0.21 & 10.03 &  5.16$\pm$3.29 &  2 \\
79983  &             1999DF9 &      cla & 6.10 &  46.38 & 0.14 & 9.82 &  36.41$\pm$5.46 &  2 \\
118379  &            1999HC12 &      cla & 8.08 &  45.30 & 0.23 & 15.35 &  3.16$\pm$7.90 &  2 \\
38083  &            1999HX11 (Rhadamanthus) &      cla (i) & 7.67 &  38.98 & 0.15 & 12.75 &  9.34$\pm$4.75 &  2 \\
  &           1999RY214 &      cla & 7.26 &  45.13 & 0.18 & 13.69 &  19.47$\pm$6.64 &  2 \\
86177  &           1999RY215 &      cla & 7.02 &  45.26 & 0.24 & 22.18 &  2.65$\pm$1.38 &  2,3 \\
  &           1999XY143 &      cla & 6.20 &  43.00 & 0.08 & 7.17 &  27.53$\pm$6.43 &  2 \\
  &           2000CJ105 &      cla & 5.99 &  44.16 & 0.11 & 11.59 &  36.53$\pm$7.18 &  2 \\
  &           2000CO105 &      cla & 5.89 &  47.05 & 0.15 & 19.27 &  21.27$\pm$8.56 &  2 \\
  &           2000CP104 &      cla & 6.86 &  44.25 & 0.10 & 9.48 &  21.43$\pm$5.53 &  2 \\
  &             2000KK4 &      cla & 6.27 &  41.26 & 0.08 & 19.13 &  26.81$\pm$7.47 &  2 \\
  &            2000OH67 &      cla & 7.49 &  44.16 & 0.02 & 5.63 &  24.60$\pm$5.30 &  3 \\
150642  &            2001CZ31 &      cla & 5.88 &  45.19 & 0.12 & 10.22 &  4.00$\pm$0.70 &  3 \\
443843  &   2001FO185 (o3e23PD) &      cla & 7.09 &  46.45 & 0.12 & 10.64 &  26.50$\pm$3.70 &  1 \\
469361  &            2001HY65 &      cla & 6.35 &  43.16 & 0.12 & 17.14 &  20.10$\pm$1.55 &  2,3 \\
  &            2001KA77 &      cla & 5.41 &  47.41 & 0.09 & 11.92 &  35.09$\pm$1.29 &  2,3 \\
  &            2001PK47 &      cla & 7.49 &  39.76 & 0.07 & 8.73 &  8.70$\pm$3.10 &  3 \\
  &           2001QC298 &      cla & 6.49 &  46.27 & 0.12 & 30.57 &  -1.25$\pm$1.45 &  2,3 \\
  &           2001QR297 &      cla & 6.77 &  44.40 & 0.03 & 5.14 &  24.60$\pm$4.10 &  3 \\
  &   2001RY143 (o4h48) &      cla & 6.80 &  42.08 & 0.15 & 6.91 &  29.50$\pm$2.10 &  1 \\
  &           2002PD155 &      cla & 7.42 &  43.23 & 0.01 & 5.77 &  18.00$\pm$5.90 &  3 \\
  &             2003LD9 &      cla & 6.92 &  47.26 & 0.17 & 6.97 &  31.30$\pm$3.30 &  3 \\
  &            2003QQ91 &      cla & 7.87 &  38.77 & 0.07 & 5.41 &  15.30$\pm$9.00 &  2 \\
307616  &            2003QW90 &      cla & 5.02 &  43.77 & 0.08 & 10.36 &  32.36$\pm$5.55 &  2 \\
  &           2003UY413 &      cla & 7.01 &  47.25 & 0.21 & 20.74 &  25.50$\pm$9.70 &  2 \\
  &           2004PA112 &      cla & 7.54 &  38.92 & 0.11 & 32.93 &  6.00$\pm$1.90 &  3 \\
504847  &   2010RE188 (o3l18) &      cla & 6.18 &  46.01 & 0.15 & 6.75 &  8.50$\pm$0.60 &  1 \\
  &   2013GG138 (o3e44) &      cla & 6.34 &  47.46 & 0.03 & 24.61 &  44.60$\pm$1.70 &  1 \\
  &   2013GO137 (o3e29) &      cla & 7.09 &  41.42 & 0.09 & 29.25 &  20.70$\pm$2.30 &  1 \\
500886  &   2013JN65 (o3o28) &      cla (i) & 7.23 &  40.67 & 0.01 & 19.64 &  7.80$\pm$0.70 &  1 \\
  &   2013JR65 (o3o21) &      cla & 7.53 &  46.20 & 0.19 & 11.71 &  0.40$\pm$2.10 &  1 \\
505448  &   2013SA100 (o3l79) &      cla & 5.75 &  46.30 & 0.17 & 8.48 &  13.10$\pm$0.48 &  1 \\
  &    2013SZ99 (o3l15) &      cla & 7.52 &  38.28 & 0.02 & 19.84 &  14.70$\pm$1.20 &  1 \\
  &   2014UH225 (o4h29) &      cla & 7.30 &  38.64 & 0.04 & 29.53 &  7.80$\pm$1.30 &  1 \\
511553  &   2014UK225 (o4h19) &      cla (i) & 7.43 &  43.53 & 0.13 & 10.69 &  31.40$\pm$1.30 &  1 \\
511554  &   2014UL225 (o4h20) &      cla & 7.24 &  46.34 & 0.20 & 7.95 &  6.40$\pm$2.10 &  1 \\
  &          2015RG277 (o5s45) &      cla & 6.79 &  42.96 & 0.01 & 12.09 &  32.80$\pm$1.20 &  1 \\
  &          2015RJ277 (o5s32) &      cla & 7.12 &  46.69 & 0.20 & 5.52 &  11.70$\pm$0.70 &  1 \\
              \hline
\multicolumn{9}{c}{ Resonant Objects } \\
  &           2006RJ103 &      1:1 & 7.67 &  30.04 & 0.03 & 8.16 &  6.16$\pm$2.69 &  2 \\
  &           2007VL305 &      1:1 & 8.42 &  30.03 & 0.06 & 28.12 &  15.77$\pm$6.10 &  2 \\
  &            1998UU43 &      4:3 & 7.24 &  36.46 & 0.13 & 9.58 &  16.40$\pm$7.23 &  2 \\
143685  &           2003SS317 &      4:3 & 7.88 &  36.46 & 0.24 & 5.91 &  31.31$\pm$3.89 &  2 \\
  &           2004TX357 &      4:3 & 8.78 &  36.58 & 0.21 & 16.26 &  11.65$\pm$2.77 &  2 \\
  &           2005ER318 &      4:3 & 7.92 &  36.52 & 0.16 & 10.42 &  11.09$\pm$2.55 &  2 \\
  &   2014UD229 (o4h13) &      4:3 & 8.18 &  36.40 & 0.15 & 6.85 &  16.40$\pm$0.80 &  1 \\
  &   2014UX228 (o4h18) &      4:3 & 7.35 &  36.36 & 0.17 & 20.66 &  7.00$\pm$1.50 &  1 \\
15789  &              1993SC &      3:2 & 7.02 &  39.52 & 0.19 & 5.16 &  39.84$\pm$8.77 &  2 \\
20108  &             1995QZ9 &      3:2 & 8.14 &  39.45 & 0.15 & 19.56 &  16.40$\pm$5.60 &  2 \\
  &            1996RR20 &      3:2 & 6.92 &  39.53 & 0.17 & 5.31 &  37.28$\pm$11.41 &  2 \\
15875  &            1996TP66 &      3:2 & 7.23 &  39.38 & 0.33 & 5.69 &  32.58$\pm$8.05 &  2 \\
118228  &            1996TQ66 &      3:2 & 7.54 &  39.41 & 0.12 & 14.66 &  36.47$\pm$11.89 &  2 \\
24952  &             1997QJ4 &      3:2 & 7.61 &  39.36 & 0.23 & 16.57 &  6.35$\pm$6.26 &  2 \\
91133  &           1998HK151 &      3:2 & 7.17 &  39.36 & 0.23 & 5.94 &  7.81$\pm$3.90 &  2 \\
91205  &            1998US43 &      3:2 & 8.06 &  39.26 & 0.13 & 10.61 &  1.32$\pm$4.33 &  2 \\
  &            1998WS31 &      3:2 & 8.21 &  39.36 & 0.20 & 6.74 &  18.76$\pm$9.68 &  2 \\
69990  &            1998WU31 &      3:2 & 8.22 &  39.23 & 0.19 & 6.58 &  11.54$\pm$5.32 &  2 \\
  &            1998WV31 &      3:2 & 7.88 &  39.25 & 0.27 & 5.72 &  17.63$\pm$8.15 &  2 \\
69986  &            1998WW24 &      3:2 & 8.30 &  39.43 & 0.23 & 13.93 &  17.08$\pm$3.99 &  2,3 \\
  &            1998WZ31 &      3:2 & 8.26 &  39.50 & 0.17 & 14.60 &  3.16$\pm$6.60 &  2 \\
47171  &            1999TC36 (Lempo) &      3:2 & 5.15 &  39.41 & 0.23 & 8.41 &  34.21$\pm$3.43 &  2 \\
  &            1999TR11 &      3:2 & 8.37 &  39.40 & 0.24 & 17.14 &  40.30$\pm$11.82 &  2 \\
  &            2000FV53 &      3:2 & 7.81 &  39.18 & 0.16 & 17.34 &  2.17$\pm$2.05 &  2,3 \\
47932  &           2000GN171 &      3:2 & 5.92 &  39.35 & 0.28 & 10.81 &  25.16$\pm$3.04 &  2,4 \\
  &             2000YH2 &      3:2 (i) & 8.05 &  39.23 & 0.30 & 12.91 &  6.00$\pm$1.50 &  3 \\
131318  &           2001FL194 &      3:2 & 7.88 &  39.31 & 0.17 & 13.70 &  6.00$\pm$1.90 &  3 \\
469362  &            2001KB77 &      3:2 & 7.52 &  39.57 & 0.28 & 17.52 &  14.03$\pm$5.02 &  2,4 \\
469372  &           2001QF298 &      3:2 & 5.25 &  39.38 & 0.11 & 22.35 &  3.82$\pm$3.30 &  2,4 \\
139775  &           2001QG298 &      3:2 & 7.31 &  39.35 & 0.19 & 6.49 &  24.60$\pm$1.10 &  3 \\
126155  &           2001YJ140 &      3:2 & 7.51 &  39.43 & 0.29 & 5.98 &  10.40$\pm$0.80 &  3 \\
55638  &            2002VE95 &      3:2 & 5.41 &  39.29 & 0.29 & 16.33 &  39.36$\pm$2.41 &  2,4 \\
 84719 &           2002VR128 &      3:2 & 5.19 &  39.42 & 0.26 & 14.01 &  26.03$\pm$1.66 &  2,4 \\
  & 2003SR317 (o3l13PD) &      3:2 & 7.98 &  39.43 & 0.17 & 8.35 &  12.00$\pm$1.10 &  1 \\
  &            2003TH58 &      3:2 & 7.17 &  39.36 & 0.09 & 27.93 &  6.33$\pm$4.21 &  2 \\
120216  &            2004EW95 &      3:2 & 6.46 &  39.38 & 0.32 & 29.28 &  3.07$\pm$0.62 &  2,3,4 \\
469708  &           2005GE187 &      3:2 & 7.51 &  39.34 & 0.33 & 18.25 &  31.37$\pm$1.97 &  2,3 \\
  &           2005TV189 &      3:2 & 7.72 &  39.41 & 0.19 & 34.39 &  9.80$\pm$1.60 &  3 \\
444745  &  2007JF43 (o3o20PD) &      3:2 & 5.27 &  39.36 & 0.18 & 15.08 &  35.10$\pm$0.60 &  1 \\
  &   2010TJ182 (o4h07) &      3:2 & 7.68 &  39.65 & 0.28 & 9.50 &  9.40$\pm$0.90 &  1 \\
  &   2013GH137 (o3e02) &      3:2 & 8.32 &  39.44 & 0.23 & 13.47 &  20.00$\pm$2.00 &  1 \\
  &   2013JB65 (o3o09) &      3:2 & 8.13 &  39.40 & 0.19 & 24.90 &  19.90$\pm$1.30 &  1 \\
  &   2013JD65  (o3o15) &      3:2 & 7.90 &  39.37 & 0.09 & 13.02 &  23.90$\pm$2.00 &  1 \\
500883  &    2013JJ65 (o3o27) &      3:2 & 7.22 &  39.37 & 0.26 & 19.82 &  40.30$\pm$1.80 &  1 \\
  &   2014UO229 (o4h11) &      3:2 & 8.25 &  39.45 & 0.16 & 10.09 &  16.06$\pm$0.57 &  1 \\
  &   2014UV228 (o4h09) &      3:2 & 8.49 &  39.49 & 0.23 & 10.13 &  13.00$\pm$1.40 &  1 \\
  &   2014UX229 (o4h05) &      3:2 & 8.04 &  39.64 & 0.34 & 15.97 &  12.20$\pm$1.00 &  1 \\
  &           1999CX131 &      5:3 & 7.17 &  42.24 & 0.23 & 9.76 &  14.40$\pm$6.08 &  2 \\
126154  &           2001YH140 &      5:3 & 5.63 &  42.33 & 0.14 & 11.08 &  14.10$\pm$1.94 &  2 \\
149349  &           2002VA131 &      5:3 & 6.72 &  42.27 & 0.24 & 7.07 &  38.16$\pm$4.07 &  2 \\
  &           2005SE278 &      5:3 & 7.06 &  42.31 & 0.11 & 6.89 &  36.22$\pm$5.72 &  2 \\
434709  &            2006CJ69 &      5:3 & 7.62 &  42.18 & 0.23 & 17.92 &  37.17$\pm$2.85 &  2 \\
469306  &           1999CD158 &      7:4 & 5.10 &  43.66 & 0.14 & 25.48 &  18.33$\pm$8.51 &  2 \\
118378  &            1999HT11 &      7:4 (i) & 7.14 &  43.74 & 0.11 & 5.05 &  32.37$\pm$4.44 &  2 \\
60620  &             2000FD8 &      7:4 & 6.71 &  43.75 & 0.22 & 19.51 &  28.97$\pm$2.29 &  2,3 \\
182222  &             2000YU1 &      7:4 (i) & 6.68 &  43.59 & 0.10 & 6.11 &  18.98$\pm$3.75 &  2 \\
  &            2003QX91 &      7:4 & 8.83 &  43.66 & 0.25 & 27.68 &  11.50$\pm$7.30 &  3 \\
  &            2004OQ15 &      7:4 & 7.22 &  43.65 & 0.12 & 9.73 &  6.71$\pm$3.62 &  2 \\
  &           2005SF278 &      7:4 & 6.56 &  43.73 & 0.19 & 13.34 &  33.59$\pm$4.89 &  2 \\
437915  &            2002GD32 &      9:5 & 6.16 &  44.48 & 0.14 & 6.59 &  34.38$\pm$5.19 &  2 \\
511555  &   2014UM225 (o4h31) &      9:5 & 7.21 &  44.48 & 0.10 & 18.30 &  22.30$\pm$1.90 &  1 \\
  &            1997SZ10 &      2:1 & 8.37 &  47.89 & 0.36 & 11.80 &  28.36$\pm$9.51 &  2 \\
26308  &           1998SM165 &      2:1 & 5.96 &  47.66 & 0.37 & 13.50 &  25.15$\pm$2.21 &  2,3 \\
137295  &           1999RB216 &      2:1 & 7.12 &  47.55 & 0.29 & 12.69 &  17.31$\pm$1.43 &  2,3 \\
130391  &            2000JG81 &      2:1 & 8.02 &  47.42 & 0.28 & 23.46 &  11.16$\pm$3.14 &  2,3 \\
  &           2004TV357 &      2:1 & 6.76 &  47.72 & 0.28 & 9.76 &  -0.62$\pm$0.80 &  2,3 \\
  &            2005CA79 &      2:1 & 5.24 &  47.78 & 0.22 & 11.67 &  15.60$\pm$0.77 &  2,3 \\
308379  &            2005RS43 &      2:1 & 5.18 &  47.80 & 0.20 & 10.01 &  7.89$\pm$0.67 &  2,3 \\
470083  &           2006SG369 &      2:1 & 7.79 &  47.78 & 0.37 & 13.59 &  28.62$\pm$3.73 &  2 \\
  &   2013GW136 (o3e05) &      2:1 & 7.42 &  47.74 & 0.34 & 6.66 &  21.20$\pm$0.60 &  1 \\
500877  &   2013JE64 (o3o18) &      2:1 & 7.94 &  47.76 & 0.28 & 8.34 &  -2.60$\pm$8.50 &  1 \\
182397  &           2001QW297 &      9:4 & 6.94 &  51.67 & 0.23 & 17.04 &  27.25$\pm$6.53 &  2 \\
181867  &           1999CV118 &      7:3 & 7.35 &  52.77 & 0.29 & 5.48 &  25.97$\pm$7.00 &  2 \\
95625  &            2002GX32 &      7:3 (i) & 7.77 &  53.07 & 0.38 & 13.94 &  37.59$\pm$10.41 &  2 \\
79978  &           1999CC158 &     12:5 (i) & 5.70 &  53.88 & 0.28 & 18.75 &  24.74$\pm$5.47 &  2 \\
119878  &           2002CY224 &     12:5 (i) & 6.18 &  53.85 & 0.35 & 15.73 &  32.65$\pm$3.96 &  2 \\
69988  &            1998WA31 &      5:2 & 7.56 &  55.18 & 0.43 & 9.46 &  17.97$\pm$11.42 &  2 \\
26375  &             1999DE9 &      5:2 & 5.03 &  55.41 & 0.42 & 7.62 &  20.45$\pm$3.32 &  2 \\
60621  &             2000FE8 &      5:2 & 6.76 &  55.29 & 0.40 & 5.87 &  12.19$\pm$1.94 &  2 \\
119068  &            2001KC77 &      5:2 & 7.10 &  55.09 & 0.36 & 12.90 &  20.80$\pm$1.20 &  2 \\
  &           2001XQ254 &      5:2 & 7.92 &  55.37 & 0.44 & 7.11 &  14.71$\pm$2.90 &  2 \\
135571  &            2002GG32 &      5:2 & 7.43 &  55.29 & 0.35 & 14.68 &  38.31$\pm$4.25 &  2 \\
143707  &           2003UY117 &      5:2 & 5.71 &  55.55 & 0.41 & 7.54 &  21.97$\pm$0.80 &  2,4 \\
  &            2004EG96 &      5:2 & 8.15 &  55.53 & 0.42 & 16.21 &  10.52$\pm$2.64 &  2 \\
  &            2004HO79 &      5:2 & 7.38 &  55.21 & 0.41 & 5.62 &  26.40$\pm$9.80 &  2 \\
  &           2004TT357 &      5:2 & 7.86 &  55.34 & 0.43 & 8.98 &  12.95$\pm$3.40 &  2 \\
  &           2005SD278 &      5:2 & 6.17 &  55.50 & 0.28 & 17.85 &  18.02$\pm$1.55 &  2 \\
  &            2009YG19 &      5:2 & 6.02 &  55.67 & 0.41 & 5.15 &  26.60$\pm$4.90 &  4 \\
  &   2013GY136 (o3e09) &      5:2 & 7.32 &  55.53 & 0.41 & 10.88 &  6.40$\pm$0.60 &  1 \\
  &   2013JK64 (o3o11) &      5:2 & 7.69 &  55.25 & 0.41 & 11.08 &  35.97$\pm$0.48 &  1 \\             
  &           2000CQ105 &     13:5 (i) & 6.27 &  56.99 & 0.39 & 19.68 &  3.73$\pm$1.69 &  2,3 \\
  &           1999RJ215 &     11:4 (i) & 7.86 &  59.27 & 0.41 & 19.71 &  13.85$\pm$1.82 &  2,3 \\
500879  &    2013JH64 (o3o34) &     11:4 & 5.60 &  59.23 & 0.38 & 13.73 &  15.60$\pm$0.80 &  1 \\ 
136120 &             2003LG7 &      3:1 & 8.58 &  62.16 & 0.48 & 20.10 &  12.99$\pm$4.66 &  2 \\
385607  &           2005EO297 &      3:1 & 7.47 &  62.70 & 0.34 & 25.04 &  13.77$\pm$4.80 &  2 \\
  &           2006QJ181 &      3:1 & 7.15 &  62.58 & 0.49 & 20.06 &  11.99$\pm$3.59 &  2 \\
126619  &           2002CX154 &     11:3 (i) & 7.51 &  71.27 & 0.47 & 15.98 &  21.27$\pm$6.51 &  2 \\
  &           2007TA418 &     15:4 (i) & 7.39 &  72.42 & 0.50 & 21.96 &  14.00$\pm$2.30 &  3 \\
184212  & 2004PB112 (o5s16PD) &     27:4 (i) & 7.39 &  107.52 & 0.67 & 15.43 &  20.30$\pm$2.50 &  1 \\
  &          2007TC434 (o4h39) &      9:1 & 7.13 &  129.63 & 0.69 & 26.47 &  11.90$\pm$1.30 &  1 \\            
148209  &           2000CR105 &     20:1 (i) & 6.44 &  221.58 & 0.80 & 22.76 &  13.22$\pm$8.40 &  2 \\
474640  &           2004VN112 &     36:1 (i) & 6.22 &  327.38 & 0.85 & 25.55 &  11.50$\pm$4.25 &  2 \\
              \hline
\multicolumn{9}{c}{ Centaurs } \\
2060  &              1977UB (Chiron) &      cen & 6.53 &  13.85 & 0.39 & 6.91 &  2.62$\pm$0.88 &  2 \\
5145  &              1992AD (Pholus) &      cen & 7.55 &  20.31 & 0.57 & 24.70 &  52.82$\pm$8.23 &  2 \\
7066  &             1993HA2 (Nessus) &      cen & 9.42 &  24.54 & 0.52 & 15.65 &  43.35$\pm$2.99 &  2,3 \\
8405  &              1995GO (Asbolus) &      cen & 9.51 &  18.06 & 0.62 & 17.62 &  13.61$\pm$2.57 &  2 \\
10199  &            1997CU26 (Chariklo) &      cen & 6.83 &  15.77 & 0.17 & 23.39 &  14.43$\pm$0.91 &  2 \\
49036  &           1998QM107 (Pelion) &      cen & 10.42 &  20.02 & 0.14 & 9.37 &  20.84$\pm$10.25 &  2 \\
52975  &            1998TF35 (Cyllarus) &      cen & 8.71 &  26.16 & 0.38 & 12.64 &  34.99$\pm$1.29 &  2,3 \\
31824  &             1999UG5 (Elatus) &      cen & 10.67 &  12.59 & 0.41 & 5.60 &  24.71$\pm$1.08 &  2,3 \\
121725  &           1999XX143 (Aphidas) &      cen & 8.83 &  17.94 & 0.46 & 6.78 &  18.70$\pm$2.50 &  3 \\
54598  &           2000QC243 (Bienor) &      cen & 7.96 &  16.50 & 0.20 & 20.74 &  8.93$\pm$4.18 &  2 \\
32532  &            2001PT13 (Thereus) &      cen & 9.58 &  10.64 & 0.20 & 20.35 &  10.31$\pm$2.43 &  2 \\
119315  &            2001SQ73 &      cen & 9.24 &  17.45 & 0.18 & 17.43 &  8.83$\pm$0.77 &  2,3 \\
427507  &             2002DH5 &      cen & 10.34 &  22.03 & 0.37 & 22.47 &  2.61$\pm$3.63 &  2 \\
55576  &            2002GB10 (Amycus) &      cen & 7.90 &  25.03 & 0.39 & 13.34 &  37.88$\pm$2.16 &  2,4 \\
83982  &             2002GO9 (Crantor) &      cen & 8.96 &  19.40 & 0.28 & 12.77 &  24.61$\pm$1.08 &  2,3 \\
95626  &            2002GZ32 &      cen & 6.88 &  23.06 & 0.22 & 15.02 &  23.40$\pm$5.41 &  2 \\
250112  &            2002KY14 &      cen & 10.19 &  12.55 & 0.31 & 19.46 &  36.28$\pm$2.23 &  2,4 \\
  &           2002PQ152 &      cen & 9.50 &  25.79 & 0.19 & 9.35 &  40.40$\pm$6.70 &  4 \\
  &            2002QX47 &      cen & 8.69 &  25.43 & 0.37 & 7.28 &  2.00$\pm$3.90 &  4 \\
120061  &             2003CO1 &      cen & 9.38 &  20.72 & 0.47 & 19.74 &  11.03$\pm$5.08 &  2 \\
136204  &             2003WL7 &      cen & 8.76 &  20.14 & 0.26 & 11.17 &  13.10$\pm$1.56 &  2,4 \\
  &            2004QQ26 &      cen & 9.91 &  22.93 & 0.15 & 21.45 &  5.50$\pm$1.90 &  3 \\
160427  &            2005RL43 &      cen & 7.99 &  24.51 & 0.04 & 12.26 &  31.30$\pm$1.30 &  3 \\
447178  &            2005RO43 &      cen & 7.16 &  28.85 & 0.52 & 35.46 &  8.84$\pm$0.78 &  3,4 \\
248835  &           2006SX368 &      cen & 9.27 &  22.08 & 0.46 & 36.30 &  11.00$\pm$2.10 &  4 \\
309139  &            2006XQ51 &      cen & 10.15 &  15.85 & 0.38 & 31.61 &  4.90$\pm$3.00 &  4 \\
187661  &            2007JG43 &      cen & 9.20 &  23.95 & 0.40 & 33.13 &  7.50$\pm$0.70 &  3 \\
341275  &           2007RG283 &      cen & 8.54 &  19.91 & 0.23 & 28.76 &  11.00$\pm$3.20 &  4 \\
  &           2007RH283 &      cen & 8.49 &  15.92 & 0.34 & 21.35 &  7.35$\pm$0.66 &  3,4 \\
281371  &            2008FC76 &      cen & 9.41 &  14.69 & 0.31 & 27.12 &  29.82$\pm$2.34 &  2,4 \\
  &              2010TH &      cen & 9.05 &  18.55 & 0.32 & 26.70 &  10.00$\pm$2.10 &  4 \\
449097  &            2012UT68 &      cen & 9.44 &  20.25 & 0.38 & 15.41 &  32.70$\pm$2.50 &  4 \\
463368  &            2012VU85 &      cen & 8.39 &  29.23 & 0.31 & 15.07 &  29.00$\pm$5.00 &  4 \\
459865  &             2013XZ8 &      cen & 9.56 &  13.43 & 0.37 & 22.53 &  8.90$\pm$2.10 &  4 \\
  &   2014UJ225 (o4h01) &      cen & 10.29 &  23.19 & 0.38 & 21.32 &  12.50$\pm$1.10 &  1 \\
  &   2015RV245 (o5s05) &      cen & 10.10 &  21.98 & 0.48 & 15.39 &  9.90$\pm$2.30 &  1 \\
              \hline
\multicolumn{9}{c}{ Scattering Disk Objects } \\
15874  &            1996TL66 &      sca & 5.34 &  83.42 & 0.58 & 23.98 &  1.56$\pm$3.89 &  2 \\
33128  &            1998BU48 &      sca & 7.18 &  33.27 & 0.38 & 14.25 &  29.03$\pm$10.54 &  2 \\
181902  &           1999RD215 &      sca & 8.11 &  122.40 & 0.69 & 25.93 &  12.70$\pm$6.30 &  2 \\
91554  &           1999RZ215 &      sca & 8.07 &  101.73 & 0.70 & 25.50 &  12.71$\pm$2.20 &  2,3 \\
29981  &            1999TD10 &      sca & 8.85 &  95.78 & 0.87 & 5.96 &  10.51$\pm$0.74 &  2,3 \\
  &           2000CQ104 &      sca (i) & 8.08 &  36.64 & 0.23 & 13.51 &  8.38$\pm$3.08 &  2 \\
60608  &           2000EE173 &      sca & 8.11 &  49.20 & 0.54 & 5.95 &  13.52$\pm$2.83 &  2,3 \\
87269  &            2000OO67 &      sca & 9.33 &  562.10 & 0.96 & 20.07 &  25.98$\pm$6.34 &  2 \\
87555  &           2000QB243 &      sca & 8.77 &  34.74 & 0.56 & 6.79 &  12.57$\pm$2.56 &  2,3 \\
82158  &           2001FP185 &      sca & 6.20 &  215.84 & 0.84 & 30.80 &  19.62$\pm$3.13 &  2 \\
82155  &           2001FZ173 &      sca (i) & 6.08 &  85.33 & 0.62 & 12.72 &  19.43$\pm$2.10 &  2 \\
  &            2001KG77 &      sca (i) & 8.57 &  61.67 & 0.45 & 15.48 &  7.90$\pm$6.30 &  2 \\
  &            2002GB32 &      sca & 7.90 &  206.81 & 0.83 & 14.19 &  19.17$\pm$1.71 &  2 \\
469442  &   2002GG166 (o3e01) &      sca & 7.73 &  34.42 & 0.59 & 7.71 &  11.10$\pm$0.60 &  1 \\          
73480  &            2002PN34 &      sca & 8.73 &  30.90 & 0.57 & 16.64 &  14.20$\pm$5.50 &  2 \\
65489  &           2003FX128 (Ceto) &      sca & 6.55 &  100.62 & 0.82 & 22.30 &  13.86$\pm$0.87 &  2,3 \\
506479  &            2003HB57 &      sca & 7.64 &  159.63 & 0.76 & 15.50 &  14.38$\pm$2.30 &  2 \\
149560  &            2003QZ91 &      sca & 8.03 &  41.51 & 0.48 & 34.82 &  12.80$\pm$2.60 &  3 \\
469750  &            2005PU21 &      sca & 6.38 &  174.64 & 0.83 & 6.17 &  32.80$\pm$2.01 &  2 \\
145474  &           2005SA278 &      sca & 6.53 &  92.28 & 0.64 & 16.27 &  8.10$\pm$1.60 &  3 \\
308933  &           2006SQ372 &      sca & 7.89 &  799.41 & 0.97 & 19.46 &  25.44$\pm$1.04 &  2,3 \\
  &           2007TG422 &      sca & 6.44 &  503.03 & 0.93 & 18.59 &  14.44$\pm$3.00 &  2 \\
  &           2007VJ305 &      sca & 6.96 &  192.12 & 0.82 & 11.98 &  15.50$\pm$2.26 &  2 \\
  &    2013JO64 (o3o14) &      sca & 8.00 &  143.31 & 0.75 & 8.58 &  8.90$\pm$1.40 &  1 \\
  &    2013UR15 (o3l01) &      sca & 10.89 &  55.82 & 0.72 & 22.25 &  13.70$\pm$2.90 &  1 \\
  &   2014UQ229 (o4h03) &      sca & 9.55 &  49.87 & 0.78 & 5.68 &  35.90$\pm$1.20 &  1 \\
  &   2015RU245 (o5t04) &      sca & 9.32 &  30.99 & 0.29 & 13.75 &  22.90$\pm$0.90 &  1 \\
  &   2015RW245 (o5s06) &      sca & 8.53 &  56.47 & 0.53 & 13.30 &  18.30$\pm$1.10  &  1 \\
              \hline
\multicolumn{9}{c}{ Detached Objects } \\
  &           1999CF119 &      det & 7.23 &  88.73 & 0.56 & 19.71 &  13.19$\pm$6.01 &  2 \\
181874  &            1999HW11 &      det (i) & 6.96 &  52.68 & 0.26 & 17.20 &  13.38$\pm$6.14 &  2 \\
40314  &            1999KR16 &      det & 5.82 &  48.76 & 0.30 & 24.82 &  38.31$\pm$1.37 &  2,3 \\
  &           2000AF255 &      det & 5.96 &  48.59 & 0.25 & 30.88 &  41.96$\pm$3.08 &  2,3 \\
60458  &           2000CM114 &      det & 7.20 &  59.49 & 0.40 & 19.70 &  14.54$\pm$2.18 &  2 \\
118702  &            2000OM67 &      det & 7.33 &  98.33 & 0.60 & 23.36 &  15.67$\pm$3.31 &  2 \\
469333  &            2000PE30 &      det & 6.09 &  54.31 & 0.34 & 18.42 &  4.07$\pm$3.64 &  2 \\
182223  &             2000YC2 &      det (i) & 7.33 &  58.47 & 0.38 & 19.88 &  0.86$\pm$6.01 &  2 \\
  &           2001FM194 &      det & 7.71 &  54.05 & 0.36 & 28.69 &  12.15$\pm$3.79 &  2 \\
  &           2001QX322 &      det (i) & 6.48 &  58.10 & 0.39 & 28.53 &  14.41$\pm$1.83 &  2,3 \\
  &           2003FZ129 &      det & 7.22 &  61.79 & 0.39 & 5.79 &  9.88$\pm$2.29 &  2 \\
  &            2004OJ14 &      det & 7.25 &  55.30 & 0.29 & 22.46 &  16.60$\pm$2.33 &  2 \\
470309  &            2007JK43 &      det* & 7.20 &  46.16 & 0.49 & 44.89 &  14.30$\pm$0.80 &  3 \\
496315  &   2013GP136 (o3e39) &      det (i) & 6.42 &  150.15 & 0.73 & 33.54 &  18.50$\pm$1.00 &  1 \\
  &      2013JL64 (o3o29) &      det (i) & 7.03 &  56.77 & 0.37 & 27.67 &  15.20$\pm$1.90 &  1 \\
  \enddata
\tablenotetext{}{Dynamical classification: cla = classical belt, sca = scattering disk, cen = centaur, x:y = resonators, where x and y indicate the specific mean motion resonance. (i) Insecure classification. 
$^*$2007JK43 is nearly on a Uranus crossing orbit and has a semimajor axis that falls within the classical belt. It is classified as `detached' because no close encounter happens during the 10~My integration of its orbit.  
Orbital elements are in the barycentric referential and were computed using the \citet{Bernstein:2000bk} orbit-fitting procedure. Dynamical classification uses the \citet{Gladman:2008tu} classification scheme. See \citet{Bannister:2016cp} for additional information.}
\tablenotetext{}{H$_{\rm r}$ are the absolute magnitudes provided in `Ref.' and converted to r-band using the Synphot/STSDAS tool. \citet{Peixinho:2015bw} do not report absolute magnitude for several objects in their dataset. We retrieved these magnitudes from the MBOSS database \citep{Hainaut:2012jb} when available, or from the original papers \citep{Boehnhardt:2002ks, Sheppard:2010cz, Benecchi:2009jo, Benecchi:2011dr} otherwise. }
\tablenotetext{}{References: 1 = Col-OSSOS (\citealt{Pike:2017gf}, \citealt{Schwamb:2018} and this work), 2 = \citet{Peixinho:2015bw} and references herein, 3: \citet{Fraser:2012cs, Fraser:2015cx}, 4: \citet{Tegler:2016ep}. }
\end{deluxetable}

\section{Effects of sample selection and color classification on our analysis}
\label{sec:stats}

We explore here the effects of sample selection and color classification on our statistical analysis. 
More specifically, we test whether our statistical results are sensitive to the magnitude and inclination cuts used to define our dataset.
We further test whether varying the spectral slope value used to divide our dataset into the gray and red color classes make any difference. 

Large TNOs are characterised by different surfaces from that of small objects (see Introduction, section~\ref{sec:intro}). 
The transition size between large and small TNO surfaces is usually proposed at ${\rm H_r}$$\sim$6 \citep{Peixinho:2008gn, Fraser:2012cs}, i.e., at a diameter of 280\,km assuming a visible albedo of 0.09 \citep{Lacerda:2014hw}. 
In this work, our statistical analysis was performed including all objects with ${\rm H_r}$$>$5 (section~\ref{sec:sample}). 
We test here whether our analysis was biased by the presence in our dataset of large 5$<$${\rm H_r}$$<$6 TNOs by repeating our analysis considering only objects with ${\rm H_r}$$>$6. 
Our results are summarised at Table~\ref{tab:stats3}. They are statistically similar to those presented in section~\ref{sec:results}. 

We further test the effects of changing the inclination cut used for the selection of our dataset. 
Our inclination cut of $i=5\degr$ should ensure minimal contamination of our sample by cold classicals (section~\ref{sec:sample}). 
Let's assume, however, that our sample comprises a significant fraction of cold classical objects up to i=6$\degr$. 
Repeating our statistical analysis only for objects located at orbital inclinations i$>$6$\degr$ reveals similar results to that of section~\ref{sec:results} (Table~\ref{tab:stats3}). 

\begin{deluxetable}{lccccccc}[h!]
\tabletypesize{ \scriptsize}
\tablecaption{ \label{tab:stats3}
Effects of magnitude and inclination cuts. }
\tablehead{\colhead{  } & \multicolumn{1}{c}{ hc } & \multicolumn{1}{c}{ res } & \multicolumn{1}{c}{ cen+sca+det } & \multicolumn{1}{c}{ all } & \multicolumn{1}{c}{ all--cen } }
\startdata
& \multicolumn{4}{c}{ ${\rm H_r}$$>$6 }\\
2-sample KS-test       &  98.0 \% &  94.1  \% &  99.3 \% &  {\bf $>$99.9} \% & {\bf 99.9} \% \\ 
2-sample AD-test       &  95.8 \% &  98.9  \% &  99.1 \% &  {\bf $>$99.9} \% &  {\bf $>$99.9} \% \\ 
T-test of means          &  99.3 \% &  99.3  \% &  99.4 \% &  {\bf $>$99.9} \% &  {\bf $>$99.9} \% \\ 
Spearman correlation &  96.0 \% &  96.8  \% &  89.7 \% &  {\bf 99.8} \%       &  99.3 \% \\ 
\hline
& \multicolumn{4}{c}{ i$>$6$\degr$ }\\
2-sample KS-test       &  98.4 \% &  87.3  \% &  96.5 \% &  {\bf $>$99.9} \% & {\bf 99.7} \% \\ 
2-sample AD-test       &  96.4 \% &  89.2  \% &  95.1 \% &  {\bf $>$99.9} \% & {\bf 99.9} \% \\ 
T-test of means          &  99.4 \% &  97.4  \% &  97.2 \% &  {\bf $>$99.9} \% & {\bf $>$99.9} \% \\ 
Spearman correlation &  92.6 \% &  89.2  \% &  77.9 \% &  98.8 \%             &  97.7 \% \\ 
\enddata
\tablenotetext{}{ Same acronyms as in Table~\ref{tab:stats1}. }
\end{deluxetable}

Finally, we test if our statistics are sensitive to the spectral slope value {\rm s$_{lim}$ used to divide our dataset into the gray and red TNO classes. 
By adopting the two most extreme spectral slope values from Table~\ref{tab:slope} (17.4 and 23.8${\rm \%/(10^3 \AA)}$), we find that our results hold for the overall dataset (Table~\ref{tab:stats4}). 
Choosing the lowest spectral slope value weakens our results for the hot classical population as one previously unclassified high-i object falls into the red class. 
Choosing the highest value weakens our results for the centaur/scattering/detached objects as a significantly higher fraction of low-i objects fall into the gray class. 

\begin{deluxetable}{lccccccc}[h!]
\tabletypesize{ \scriptsize}
\tablecaption{ \label{tab:stats4}
Effects of color classification. }
\tablehead{\colhead{  } & \multicolumn{1}{c}{ hc } & \multicolumn{1}{c}{ res } & \multicolumn{1}{c}{ cen+sca+det } & \multicolumn{1}{c}{ all } & \multicolumn{1}{c}{ all--cen } }
\startdata
& \multicolumn{4}{c}{ {\rm s$_{lim}$}$=$17.4${\rm \%/(10^3 \AA)}$ }\\
2-sample KS-test &  65.8 \% &  92.8  \% &  97.8 \% &  {\bf $>$99.9} \% &  99.5 \% \\ 
2-sample AD-test &  70.1 \% &  97.2  \% &  97.1 \% &  {\bf $>$99.9} \% &  99.5 \% \\ 
T-test of means    &  93.2 \% &  98.5  \% &  98.1 \% &  {\bf $>$99.9} \% &  {\bf 99.8} \%\\ 
\hline
& \multicolumn{4}{c}{ {\rm s$_{lim}$}$=$23.8${\rm \%/(10^3 \AA)}$ }\\
2-sample KS-test &  99.3 \% &  87.5  \% &  78.5 \% &  {\bf 99.9} \%       & 98.6 \% \\ 
2-sample AD-test &  94.5 \% &  96.7  \% &  72.7 \% &  {\bf 99.8} \%       & {\bf 99.8} \% \\ 
T-test of means    &  99.0 \% &  98.2  \% &  90.6 \% &  {\bf $>$99.9} \% & {\bf 99.9} \% \\ 
\enddata
\tablenotetext{}{ Same acronyms as in Table~\ref{tab:stats1}. }
\end{deluxetable}

\end{document}